\documentclass{aa}
\usepackage[varg]{txfonts}
\usepackage{booktabs,graphicx,siunitx,xspace,natbib}
\usepackage{makecell,longtable}
\DeclareSIUnit \parsec {pc}
\DeclareSIUnit \pc     {\parsec}
\DeclareSIUnit \kpc    {\kilo \parsec}
\DeclareSIUnit \msun   {\ensuremath{M_\sun}}
\DeclareSIUnit \year   {a}
\newcommand{\rpp}{Rep.Prog.Phys.}%
         
\begin{document}

\title{INTEGRAL/SPI $\gamma$-ray line spectroscopy}
\subtitle{Response and background characteristics}

\author{
  Roland Diehl$^1$\thanks{\email{rod@mpe.mpg.de}}\and      
  Thomas Siegert$^1$\and 
  Jochen Greiner$^1$ \and
  Martin Krause$^3$ \and
  Karsten Kretschmer$^2$ \and
  Michael Lang$^1$ \and \\
  Moritz Pleintinger$^1$ \and
  Andrew W. Strong$^1$ \and
  Christoph Weinberger$^1$ \and
  Xiaoling Zhang$^1$
   } 

\institute{
  Max-Planck-Institut f\"ur extraterrestrische Physik, Giessenbachstr. 1, 
  D-85741 Garching, Germany
  \label{inst:mpe} \and
  AstroParticule et Cosmologie, Universit\`e Paris Diderot, CNRS/IN2P3, CEA, 75205 Paris Cedex 13, France 
 \label{inst:cea} \and
  University of Hertfortshire, School of Physics, Hatfield, AL10 9AB, United Kingdom
  \label{inst:hertfordshire}
  }

\date{Received 22 Aug 2017 / Accepted 24 Oct 2017}

\abstract
{ESA's space based $\gamma$-ray observatory INTEGRAL includes the spectrometer instrument 'SPI'. This is a coded mask telescope featuring a 19-element Germanium detector array  {for high-resolution gamma-ray spectroscopy}, encapsulated in a scintillation detector assembly { that provide a veto for background from charged particles}.
In space, cosmic rays irradiate spacecraft and instrument, which, { in spite of the vetoing detectors,} results in a large instrumental background from activation of those materials, and { leads to }deterioration of the
charge collection properties of the Ge detectors.}
{We aim to determine the measurement characteristics of our detectors and their evolution with time, i.e. their spectral response and instrumental background. These incur systematic variations in the SPI signal from celestial photons, hence their determination from a broad empirical database enables a reduction of underlying systematics in data analysis. For this, we explore compromises balancing temporal and spectral resolution within statistical limitations. Our goal is to enable modelling of background applicable to spectroscopic studies of the sky, accounting separately for changes of the spectral response and of instrumental background. }
{We use 13.5 years of INTEGRAL/SPI data, which consist of spectra for each detector and for each pointing of the satellite. Spectral fits to each such spectrum, with  independent but coherent treatment of continuum and line backgrounds, provides us with details about separated background components. From the strongest background lines, we first determine how the spectral response changes with time. Applying symmetry and long-term stability tests, we eliminate degeneracies and reduce statistical fluctuations of background parameters, towards a self-consistent description of the spectral response for each individual detector. Accounting for this, we then determine how the instrumental background components change in intensities and other characteristics, most-importantly their relative distribution among detectors. }
{Spectral resolution of Ge detectors in space degrades with time, up to 15\% within half a year, consistently for all detectors, and across the SPI energy range. Semi-annual annealing operations recover these losses, yet there is a small long-term degradation. The intensity of instrumental background varies anti-correlated to solar activity, in general. There are significant differences among different lines and with respect to continuum. Background lines are found to have a characteristic, well-defined and long-term consistent intensity ratio among detectors. We use this to categorise lines in groups of similar behaviour. 
The dataset of spectral-response and background parameters as fitted across the INTEGRAL mission allows studies of SPI spectral response and background behaviour in a broad perspective, and efficiently supports precision modelling of instrumental background.
 }
{}


\maketitle

%

\begin{figure}
  \centering
  \includegraphics[width=0.3\linewidth]{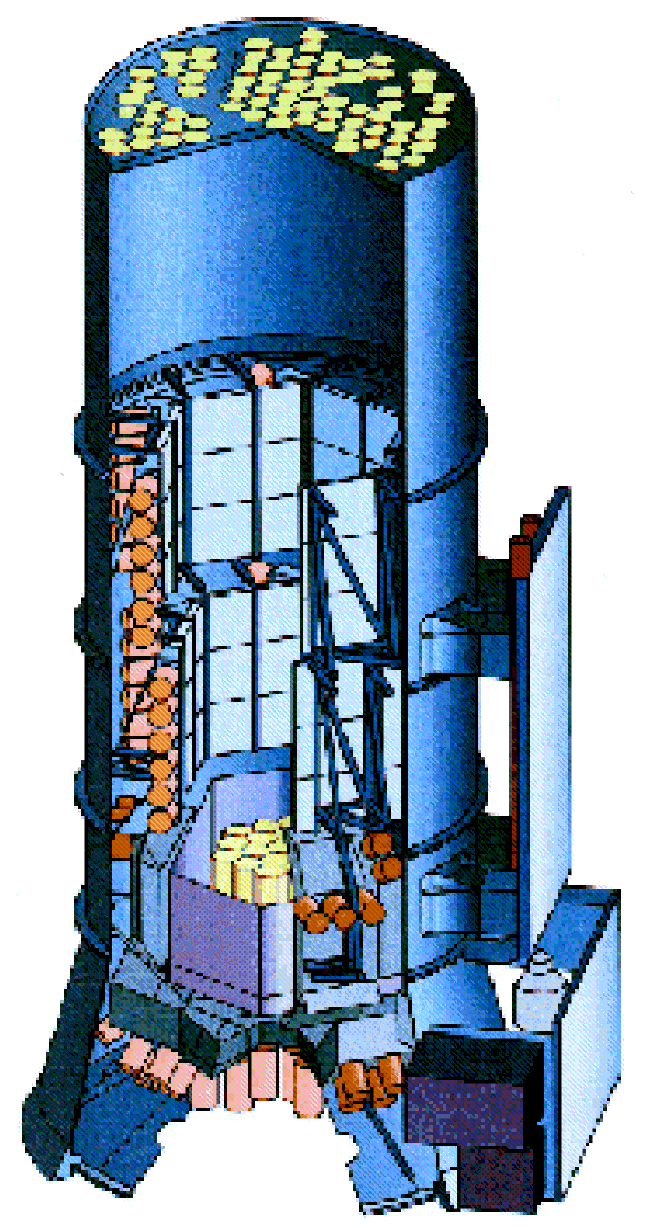}
   \includegraphics[width=0.6\linewidth]{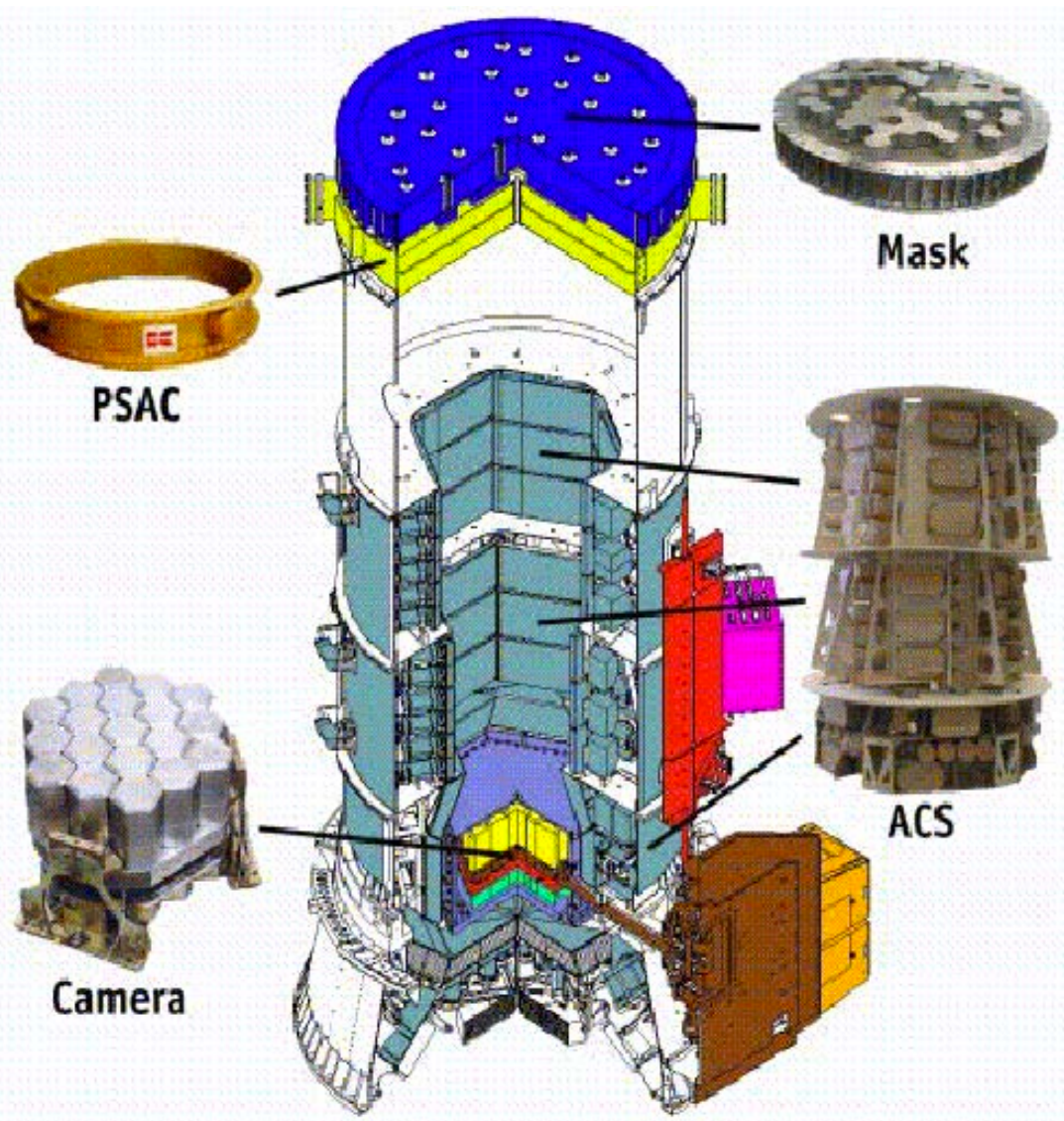}
   \caption{SPI instrument on INTEGRAL, schematic \emph{(left)} and as a cut-away showing components more clearly \emph{(right)}. Its key elements are the 19 Ge detector \emph{camera} mounted in a cryostat for operations at $\sim$80K, a \emph{mask} with tungsten elements blocking parts of the field of view for about half of the detectors, and an anti-coincidence detector system \emph{(ACS)} enclosing the entire instrument and vetoing events from prompt cosmic-ray interactions,  using BGO scintillators and a plastic scintillator plate \emph{(PSAC)}, each with photomultiplier tubes for measuring the scintillation light.}
  \label{fig:SPI-instrument}
\end{figure}

\section{Introduction}
\label{sec:intro}
 The International Gamma Ray Astrophysics Laboratory (INTEGRAL, \citet{Winkler:2003}) was launched into space by ESA in 2002 for a mission of 3+2 year nominal duration\footnote{The INTEGRAL mission has been extended after reviews every two years since; INTEGRAL may continue to operate until its controlled re-entry into the Earth atmosphere in  2029, although solar-array degradation may become critical in the mid-20$^{ies}$.}. Since then, INTEGRAL has been accumulating a legacy database of the sky in hard X-rays and soft $\gamma$-rays from eventually two decades of observations. 
The SPI coded mask telescope (Fig.~\ref{fig:SPI-instrument}) is one of the two main instruments on INTEGRAL. This instrument  has been designed as a spectrometer for $\gamma$-ray lines from celestial sources in the energy band from 20 to 8000 keV. It is described in detail in \citet{Vedrenne:2003}, pre-launch and initial in-flight calibrations are described in \citet{Attie:2003} and \citet{Roques:2003}, respectively, and five years of spacecraft and instrument operations are reviewed in \citet{Fahmy:2008}. 
Main scientific goals of the INTEGRAL space mission are  \citep{Winkler:2003,Winkler:2011}:  A survey of the high energy sky with particular attention to sources such as binary systems, active galaxies, and transients such as stellar explosions and outbursts, flares, and state transitions, and the study of cosmic nucleosynthesis using nuclear lines. The latter is a unique domain of SPI due to its spectral range and resolution  
\citep[see review by][]{Diehl:2013}. 

Gamma-ray spectroscopy presents its unique challenges: 
Data from observations over several years need to be analysed coherently for typical  source intensities of 10$^{-5}$ph~cm$^{-2}$s$^{-1}$. In a 10$^6$~second observation, such a typical source would yield about 2500 source counts within a total of $\sim$3~10$^7$  SPI event counts. Variations of instrumental backgrounds and detector responses are significant. Therefore cosmic signals are best recognised when knowledge about instrumental backgrounds and detector responses and their variations are integrated parts of the data analysis, in an iterative forward-folding approach.

Here we describe ingredients of methods towards high-resolution $\gamma$-ray spectroscopy with SPI, as  obtained from many mission years, and the collective studies performed on a variety of sources of cosmic $\gamma$-rays. In particular, we show how we exploit the characteristic signatures in the instrumental background to extract details about the spectral response. These provide a better understanding of the background itself, { and enable an } improved inclusion of the instrumental measurement- and background-characteristics in specific astrophysical studies.  

This paper is organised as follows: In Section \ref{sec:spi-analysis}, we set the context and first describe the general approach of analysing SPI data, defining terminologies and the essential data properties. Section \ref{sec:mission-analysis}  presents our method to extract spectral-response and background detail for the multi-year data archive. In Sections \ref{sec:spec-response} and \ref{sec:background}, we then describe the results obtained for the spectral response, and instrumental background, respectively. 
An Appendix provides details on background lines and on the database of spectral-fit results.

\begin{figure}
  \centering
  \includegraphics[width=\linewidth]{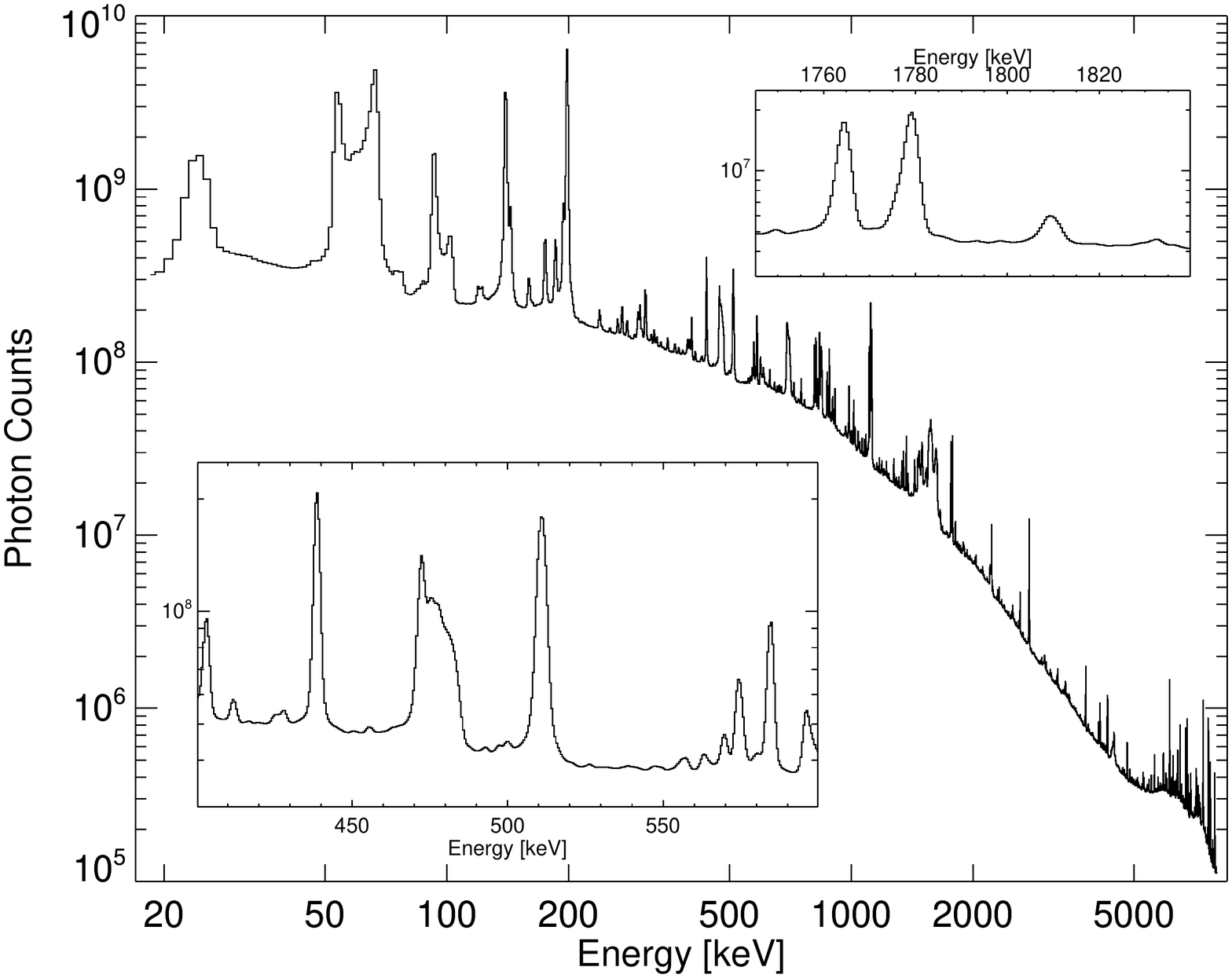} 
    \includegraphics[width=\linewidth]{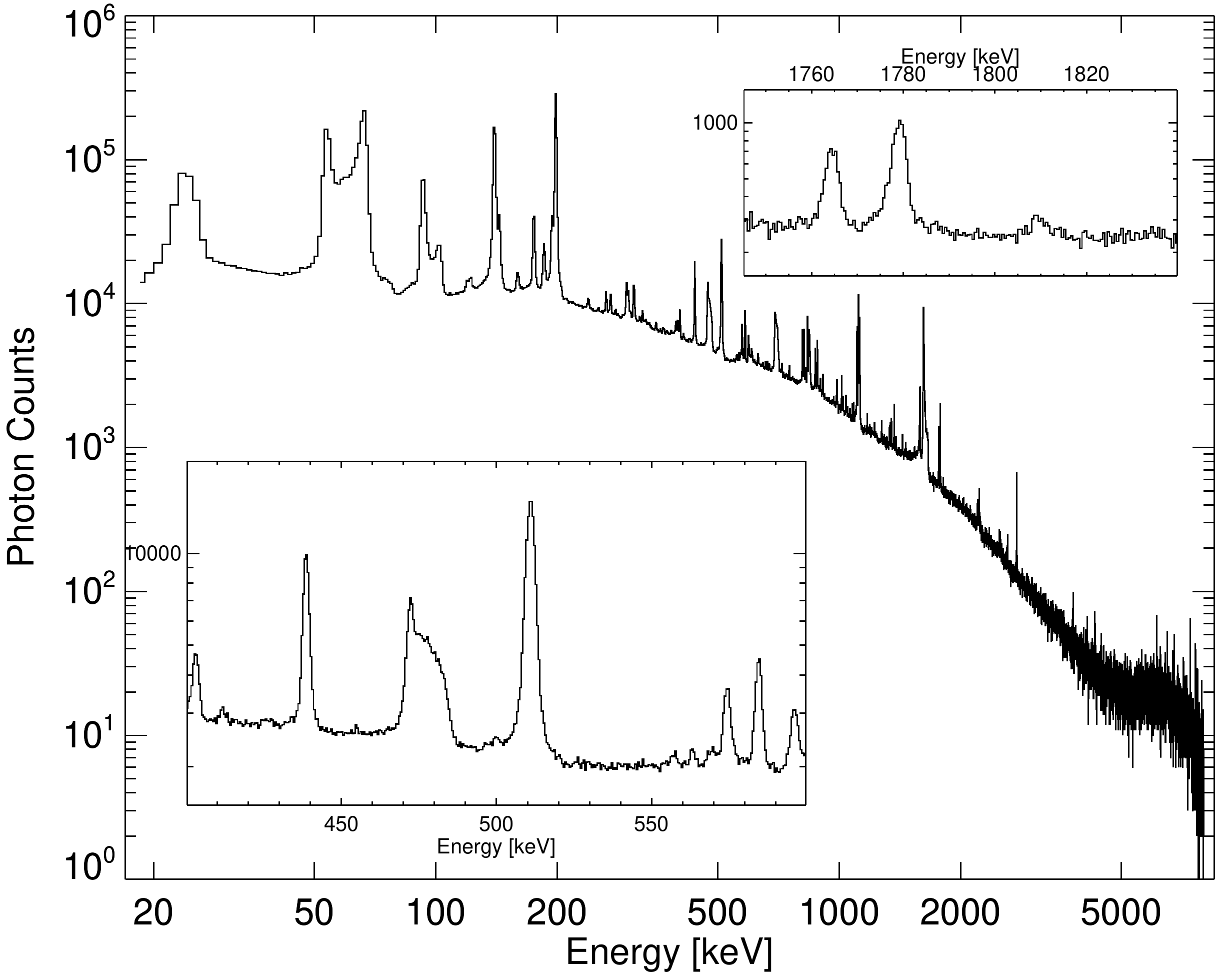} 
      \includegraphics[width=\linewidth]{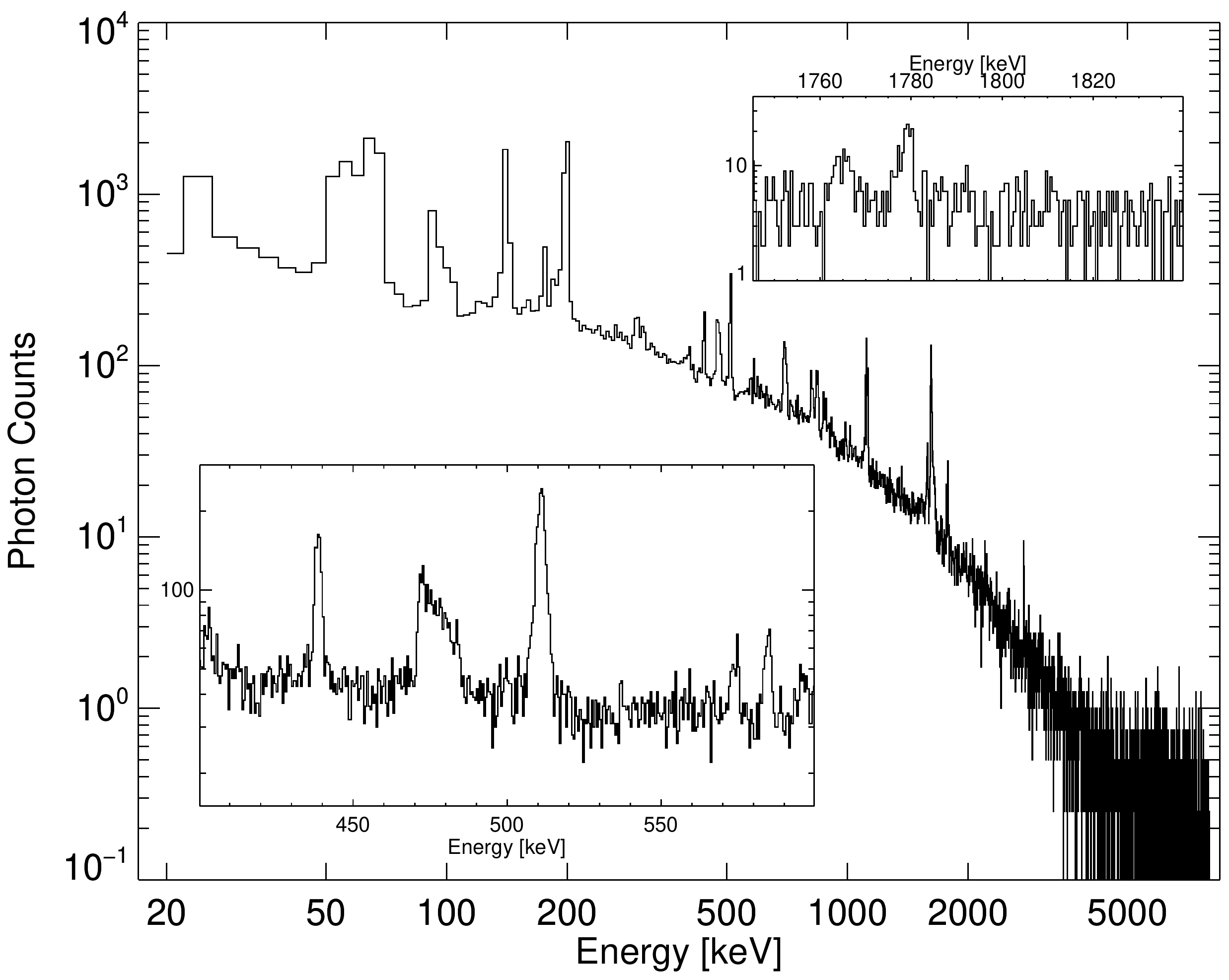} 
   \caption{Energy spectra measured with SPI during the INTEGRAL mission. {\it Above:} Mission-integrated all-detector spectrum (orbits 43-1730). This allows to identify even weakest background features. Inserts show spectral regions near the two brightest celestial lines, the 511 keV line from positron annihilation ({\it lower-left}), and the $^{26}$Al line at 1809~keV from diffuse nucleosynthesis throughout the Galaxy ({\it upper-right}). {\it Middle and below:} The same spectrum, now accumulated just for detector no. 0 and orbit no. 1000 {\it (middle)} , and for one detector (no. 0) and one pointing {\it (below)}. 
   }
  \label{fig:SPI-spectrum_all-mission}
\end{figure}
%
\section{Analysis of SPI data} 
\label{sec:spi-analysis}
%

\begin{figure}
  \centering
  \includegraphics[width=0.8\linewidth]{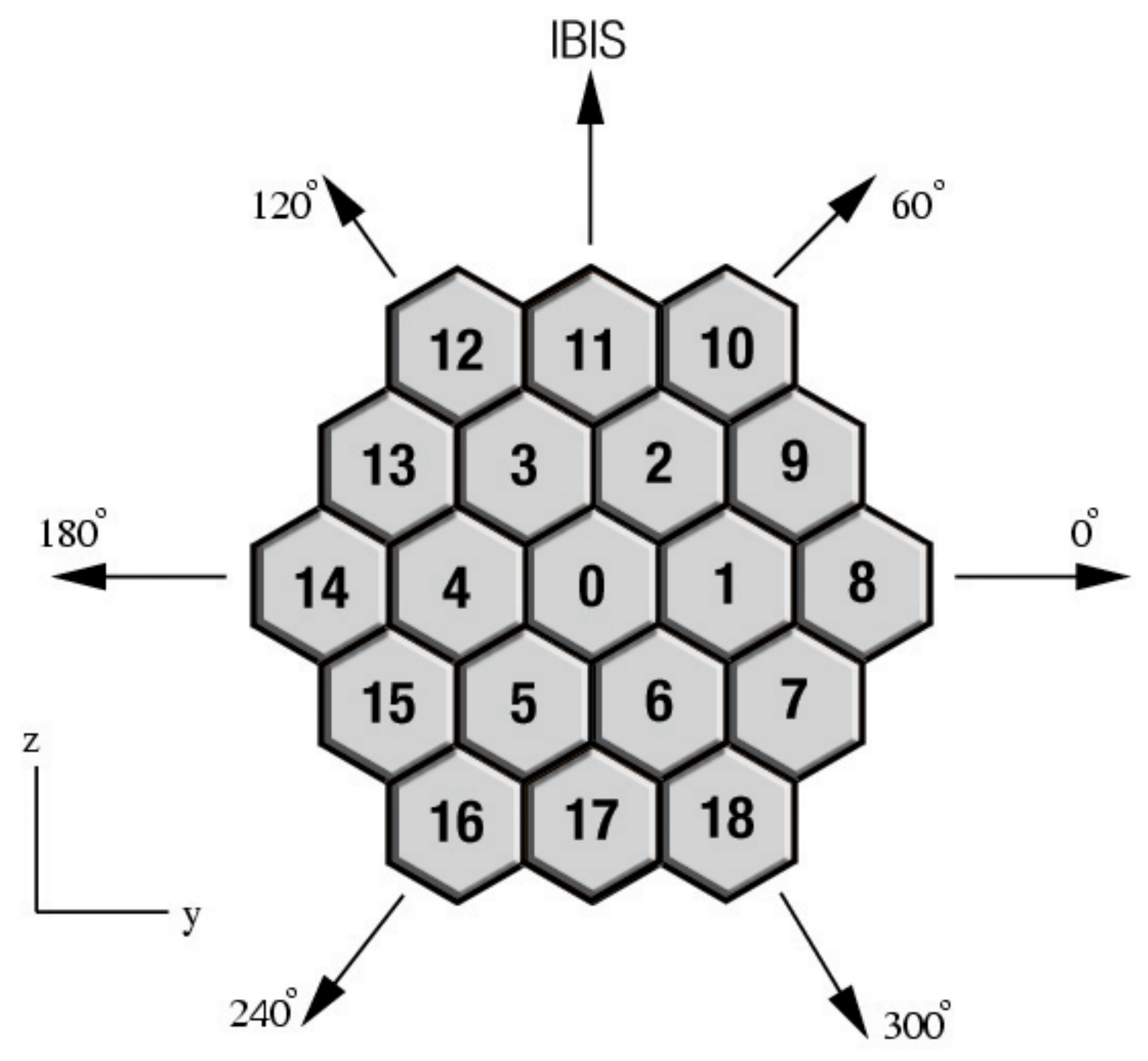}
    \includegraphics[width=0.99\linewidth]{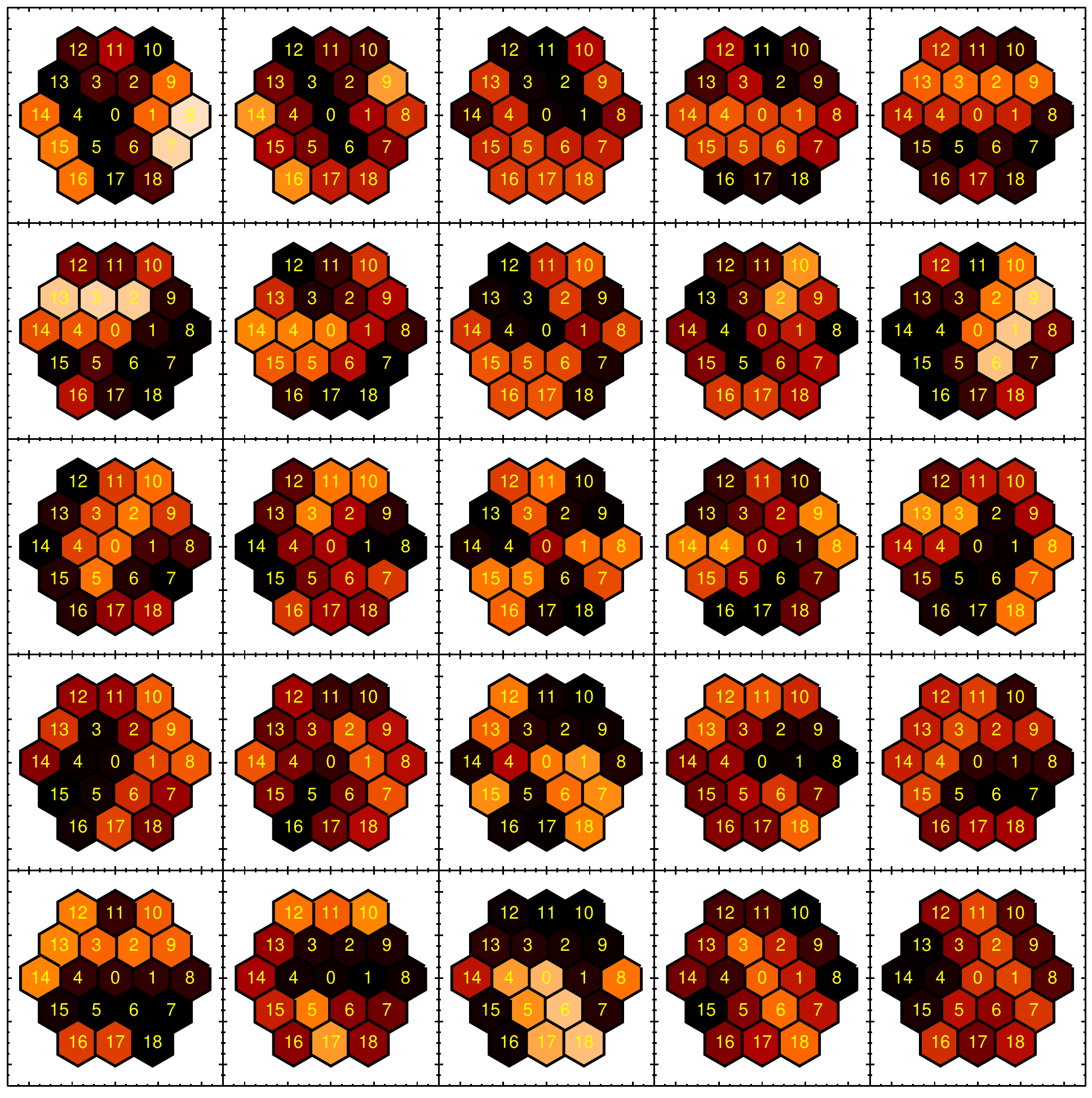}
   \caption{ { SPI Ge detector arrangement and shadograms from the mask.} The 19 Ge detectors of the SPI camera are arranged in a hexagonal, densely packed, configuration \emph{(top)}. Detectors are numbered inside-out, starting from 0 for the central detector. { Shadowgrams are }cast by the mask onto the Ge camera. The figure set \emph{below}  shows { these shadowgrams} for a point source on axis, and for the 25  telescope reorientations in steps of 2.1 degrees (pointings) that comprise a standard 5x5 dither pattern. Intensity received by each detector is colour-coded from black (none; shadowed) through red, yellow to white (full exposure).}
  \label{fig:SPI-maskpattern}
\end{figure}

Here we describe how response and background information enters into the astrophysical analysis of SPI data, in order to illustrate critical dependencies and impacts.

\subsection{SPI primary data} 
\label{sec:dataspace}

The original data of the SPI instrument consist of \emph{events} that trigger the Ge detectors. Registration of SPI events is disabled when the anticoincidence system signals a trigger (window width 725 ns), e.g. from a charged particle interaction. 
The signals measured by front-end electronics for each event are: the time of the trigger, the detector identifier, and the signal pulse height. For events that triggered more than one of the Ge detectors within the coincidence time window of the front end electronics (350 ns), multiple time tags, detector identifiers, and pulse heights are included in a corresponding event message (\emph{Multiple Events, ME}). Pulse shape information is captured from a dedicated electronics, and added to help data selections. 
The logical scheme of the onboard electronics and how it transforms detector signals into telemetry components is illustrated in  \citet{Vedrenne:2003} (their Fig.~2 and Table 1), and described in more detail in Section~7 of \citet{Roques:2003}. 
For simplicity, here we describe only events that trigger a single Ge detector (\emph{Single Events, SE}) and pass the event selections\footnote{Event selections include pulse shape information, which was found to improve the sensitivity in the energy range 500-1100~keV; those selected single events are also called 'PE'.}.  

Pre-processing of these event data at INTEGRAL's Science Data Centre (ISDC) includes a first-order energy calibration of the pulse height to keV units, and sorting them into count histograms, for each Ge detector separately, and for  each spacecraft pointing; \emph{pointings} comprise $\sim$half-hour long intervals where the SPI telescope was oriented with its central field of view (z axis) pointing towards a particular sky direction.
Auxiliary data such as detector dead times, and the start and end times, as well as direction coordinates of each pointing complement the pre-processed database.
Fig.~\ref{fig:SPI-spectrum_all-mission} shows examples of SPI data histograms for different accumulation times; these are discussed in more detail below.

For convenience of data management, but also due to expected background variations, we group data by spacecraft \emph{orbits}\footnote{Orbits are also often called \emph{'revolutions'} in the INTEGRAL community.}. INTEGRAL's orbit is highly eccentric to include long times beyond the radiation belts and thus in more-stable background environments, with an initial perigee/apogee of 9000 km and 154,600 km, respectively. One spacecraft orbit is completed within 72 hours (three days\footnote{After ESA's orbit adjustment in January 2015 for a controlled re-entry in February 2029, the current orbit duration is 2 days and 16 hours.}). 
Additional, natural, long-term groupings of data are incurred by the regular annealing operations, and by failures of Ge detectors. 

We describe these SPI data in the form of histogram sets as event counts per data space bin, 
$d_{i,j,k}$, with $i,j,k$ being the indices of pointing, detector, and energy bin that span the data space. 
The 19 \emph{detectors} are identified by number (see Fig.~\ref{fig:SPI-maskpattern}), and \emph{energy} bins are 0.5~keV wide and span a range of 20--2000~keV in our case\footnote{The SPI energy range is subdivided into a \emph{low, 20-2000~keV} and \emph{high, 2000-8000~keV} part, according to different resolutions of the analogue-to-digital conversion electronics; we focus here on the low range, and will address the high range in a separate paper.}.
 
These data are the result of the instrument's response to the $\gamma$-ray sky and the underlying instrumental background.
Modelling the measurement process, we may write the structure of data as:
\begin{equation}
d_{i,j,k} = \sum_l R_{l;ijk} \sum_{n = 1}^{N_\mathrm{s}} \theta_n S_{nl} +
\sum_{n = N_\mathrm{s} + 1}^{N_\mathrm{s} + N_\mathrm{b}} \theta_n
B_{n;ijk}\label{eq:model-fit}
\end{equation}
Here, we identify sky model components $S_{n}$, such as point sources and diffuse emissions. These are formulated in \emph{image space}, as photon source intensities per sky direction $l$.  The instrument response matrix $R_{l;ijk}$ must be applied, to link the source locations on the sky to data, combining coordinate transformations per pointing to aspect angles, and then accounting for mask/detector configurations.
This sky signal is superimposed onto a large instrumental background, and we distinguish components e.g. from continuum and from lines reflecting specific processes. The background models are formulated in the same data space of detectors and their counts; no specific instrumental-response application is required, in particular no shadowing by the mask occurs, as background is recorded by the active detector volumes from all directions.
The scientific analysis of SPI data is typically approached as a comparison of the data as measured to predictions from models\footnote{The background cannot be simply 'subtracted'; e.g. Poissonian fluctuations might lead to \emph{negative source intensities}. A direct deconvolution also is not possible, because the response matrix is singular and inversion thus ill-defined.}.

\subsection{Astronomical response} 

The \emph{imaging information} that characterises a set of photon events coming from a celestial source is their shadowgram, i.e. how they are characteristically distributed over the 19 detectors of the camera. 
For a source shining through the mask of the SPI telescope's aperture, some of the 19 detectors are \emph{shadowed} and thus do not record events from the celestial source, while others are exposed to the source (Fig.~\ref{fig:SPI-maskpattern}). There are intermediate cases of partial shadowing, depending on the source aspect angle. 

Typically, the INTEGRAL telescope orientation is systematically varied in small steps around the target direction. This is called \emph{dithering}, and it produces additional modulation of the \emph{imaging information}  (see lower part of Fig.~\ref{fig:SPI-maskpattern}). 
The dithering pattern of successive such pointings is normally a 5$\times$5 grid around the target direction, with steps of 2.1~degrees, shown in Fig.~\ref{fig:SPI-maskpattern}; sometimes, a hexagonal pointing pattern is used, which avoids partial shadowing of detectors for point sources.

Over the typically long exposure time with hundreds to many thousands of pointings thus a well-defined modulation of the celestial signal is produced. 
This 'detector pattern' of relative counts among detectors is the key to distinguish celestial from instrumental-background events, which should not vary in this way from pointing to pointing (see below). 
For source intensities which are typical for astrophysical $\gamma$-ray lines at MeV energies of $\sim$10$^{-5}$ph~cm$^{-2}$s$^{-1}$ (see e.g. Diehl 2013), 
the 255 cm$^2$ of typically-exposed Ge detector area will capture only few photon events  per telescope pointing interval, $\sim 0.5$ events per detector for a 1-hour pointing\footnote{Estimated counts per detector are $\eta A_{camera} T_{pntg} F_{source} n_{det}^{-1} \sim (255/19)*3600*10^{-5} \sim 0.5$,  where the detection efficiency $\eta$ is close to 1 for low energies around  100 keV, and about 0.1 at energies of MeV.}. 
Therefore this modulation is realised only in a statistical way through the number of totally detected source photons. 
The response matrix $R_{l;ijk}$ encodes the \emph{imaging response}, i.e. how the occultation by the coded mask affects visibility of the source direction from each of the Ge detectors of the camera during the observations. 
Coordinate transformation between the instrument coordinate system (see top of Fig.~\ref{fig:SPI-maskpattern}) and celestial coordinates are part of this response matrix application.

The instrument's  \emph{spectral response} is not employed in this \emph{high resolution spectroscopy} analysis that aims for astrophysical $\gamma$-ray lines. The celestial photon flux is derived in spectral resolution bins finer than instrumental resolution. Any spectral analysis across wider spectral bands may apply the spectral response in a further deconvolution step.

\subsection{Instrumental background} 

\begin{figure}
  \centering
  \includegraphics[width=0.7\linewidth]{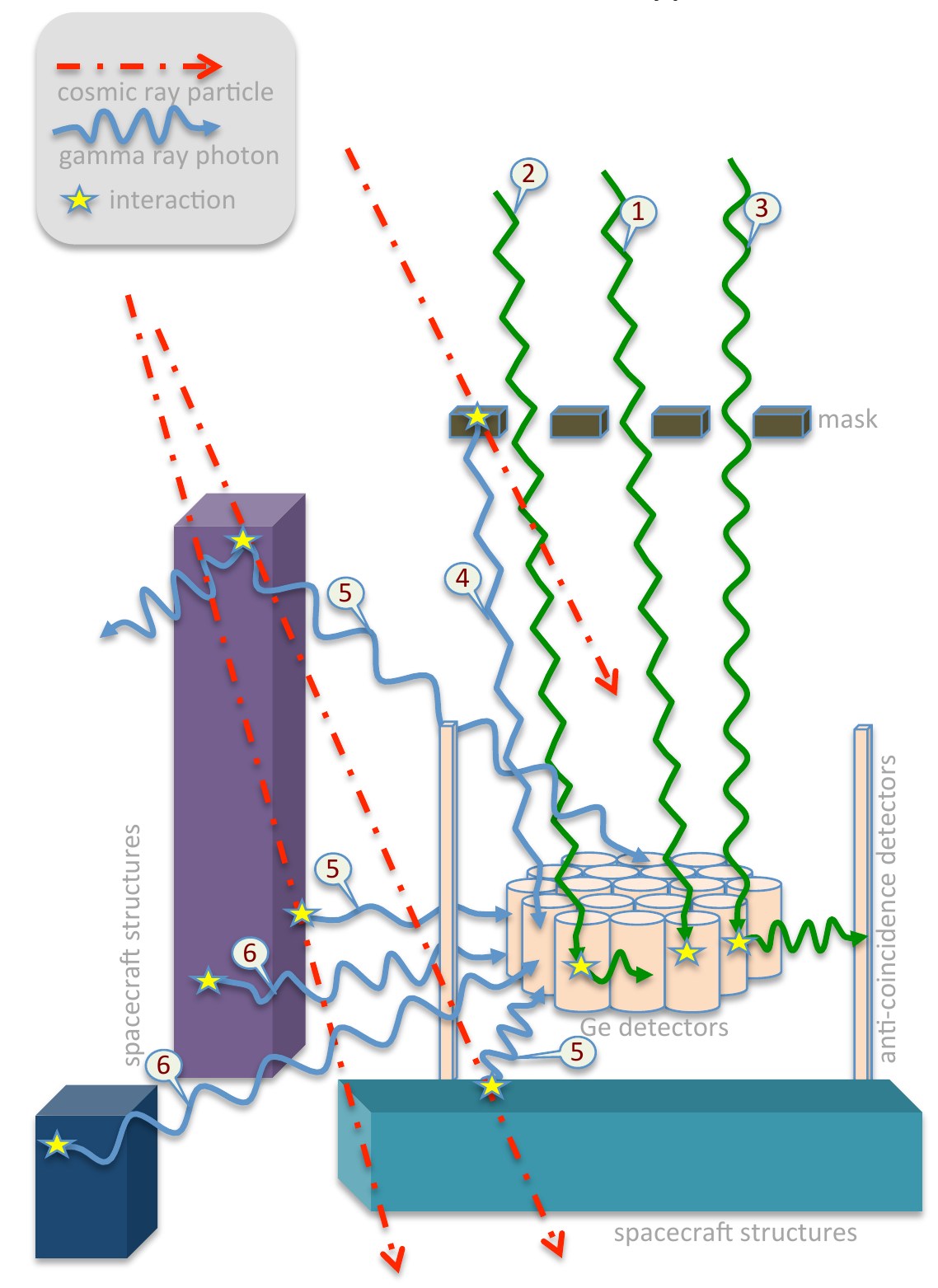}
   \caption{ { Different types of events }recorded by SPI detectors. From the three cosmic photons \emph{(green, 1-3)}, single events \emph{(SE,1)}, multiple events \emph{(ME,2)}, or self-vetoed events \emph{(3)} may arise.
   Background event types are from photons \emph{blue} which may be prompt cosmic-ray \emph{(red dashed)} related \emph{(4,5)}, or delayed and radioactivity events \emph{(6)}; some of these may, or may not, be vetoed by the anticoincidence. Structural elements with different materials may create continuum photons and lines at different, characteristic, energies.}
  \label{fig:SPI-eventTypes}
\end{figure}

SPI Ge detector spectra are dominated by instrumental background, and characterised by a continuum falling towards higher energies, with many lines from nuclear de-excitation superimposed (Fig.~\ref{fig:SPI-spectrum_all-mission}).
{ Energetic particles in general are the cause of instrumental background; for our purpose, we do not distinguish their origins from cosmic rays, or Earth albedo, or radiation belts, as we pursue an empirical characterisation rather than a bottom-up physical model\footnote{Our studies based on GEANT physics simulations, although generally providing a good guide to understand instrumental backgrounds \citep{Weidenspointner:2003}, are quantitatively not accurate enough for proper background descriptions as needed for astrophysical data analysis. However, the implications of the background created within SPI detectors for solar activity, earth magnetosphere, earth albedo, and radiation belts, are all interesting scientific questions that should/will be addressed; they are considered beyond the scope of this paper, however.}.} 
As { such} high energy particles (energies MeV to GeV) collide with atoms and nuclei of spacecraft materials, prompt and delayed radiation processes are stimulated (see illustrations in Fig.~\ref{fig:SPI-eventTypes}).  Many of these produce photons in the energy domain where SPI aims to measure cosmic $\gamma$-rays. Bremsstrahlung and electromagnetic nuclear transitions provide a \emph{prompt} background of photons, and nuclear reactions of cosmic-ray protons create secondary protons, neutrons, and nuclei, as well as excited nuclei and radioactive isotopes, which may lead to \emph{delayed} background events.

The continuum part of background originates in a variety of processes, dominated by bremsstrahlung from secondary particles, with full or partial energy deposits in detectors. 
The well-defined narrow background lines have more specific origins, each one mostly related to a single specific isotope and its nuclear de-excitation emission. The primary activation, or excitation, of a nuclear energy level above the ground state can be caused by cosmic-ray interactions, and is a resonant process, with a maximum of its cross section near the nucleus excitation energies. These are in the range from about 50 keV to 10 MeV, depending on the atomic nucleus, so that cosmic rays in the range of several tens of MeV per nucleon are most relevant. Excited nuclei also can result as daughter products from radioactive decays of unstable atomic nuclei. 

At energies below a few hundred MeV per nucleon, the near-earth spectrum (and composition) of cosmic rays is variable and not predictable in detail \citep{Potgieter:2013,Benhabiles-Mezhoud:2013}. Therefore, absolute background estimations, from simulations, or from analytical estimates, are highly uncertain; we follow an entirely-empirical approach in our work, for this reason.

During the INTEGRAL mission, the cosmic-ray intensity varies on several time scales: On the annual scale this is due to solar particle flux variations { caused by the 11-year solar activity cycle, and by }shielding variations from varying magnetic fields in the heliosphere; the latter is driven by solar activity.  On the orbital time scale, the passage near radiation belts at perigee leads to a high dose of charged particles, and this activation of spacecraft materials decays during the remainder of the orbit as the spacecraft is in quieter regions. Solar flares lead to impulsive irradiation of the spacecraft with charged particles and neutrons; during the active phases of the Sun, these may occur at frequencies of several per day. 

\begin{figure}
  \centering
  \includegraphics[width=\linewidth]{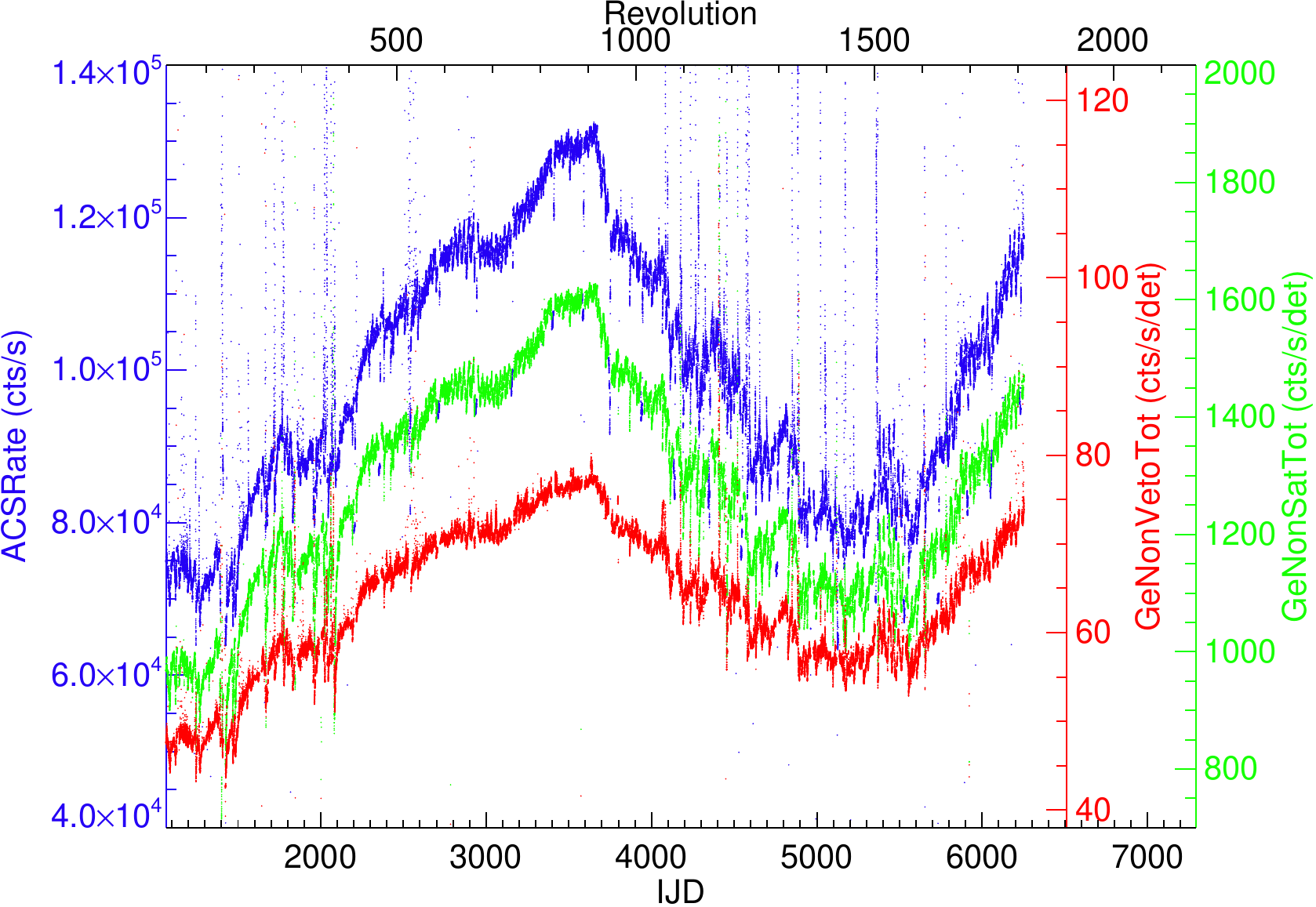}
   \caption{Evolution of SPI detector count rates over the mission. The SPI anticoincidence system mainly records charged-particle interactions, thus creating a veto signal for the Ge detector electronics; this avoids full processing of Ge detector signals for events that are coincident with a cosmic-ray hit. Shown are measured rates of the anticoincidence system (\emph{blue, left axis}), and two different event rates in the Ge detectors (\emph{green for total, red for non-vetoed and thus either delayed background or potentially celestial triggers; right-side axes}).
   Sporadic high rates are caused by solar flares.  }
  \label{fig:SPI-bgd-rates}
\end{figure}

Fig.~\ref{fig:SPI-bgd-rates} shows the charged particle flux that is measured with SPI's anticoincidence detector system (axis on the lefthand side), as well as the rate of Ge detector events,
which result from the background that is not suppressed by the anticoincidence veto signals directly (right-sided axis). 
Here  the first and last 10\% of each orbit, i.e. close to perigee and radiation belt passages, were not included; this is similar to the standard event selections used for scientific analyses of SPI data.
Nevertheless, the strong impact from solar flares is clearly seen, beyond the main long-term modulation caused by the 11-year variation of solar activity.

\subsection{Response and background aspects of SPI data analysis} 
\label{sec:dataAnalysis}

According to Equ.~\ref{eq:model-fit}, the measured data can be represented by a combination of the sky and background models, optimising their amplitude parameters through a maximum-likelihood method.
We determine the best fit for  the intensity scaling factors of a sky
intensity distribution model, together with the scaling factors for the model of the instrumental background, across an energy range of interest;  
this is the differential intensity spectrum of the sky.

The models used herein are based on prior knowledge about the sky and about instrumental background. Degeneracies may occur, and compromise the result for celestial emission.
Differences in the detector ratios and the temporal variations among those two model components are the main leverage to distinguish sky and background.   
The empirical determination of the detector pattern for backgrounds, using the data from all observations during the mission, promises to improve precision. 

In summary, our analysis approach is based on two key assumptions about instrument-specific aspects: 
\begin{itemize}
\item[(1)] 
 The relative contributions of detectors to the celestial-source signal follows the mask shadowing of detectors. 
\item[(2)] 
The instrumental background variations are not related to small spacecraft re-pointings, and retain characteristic signatures such as spectral shape and relative detector ratios across such re-pointings.
\end{itemize}
 
\noindent The second assumption allows a separation of background variations into \emph{intensity} and \emph{signature}, towards the following two parts:
\begin{itemize}
\item[(a)] 
The ratios of detector count rates are characteristic, and different, for each background line and for continuum. Hence such decomposition is required and justified. 
Characteristic detector ratio patterns can be determined from data that are largely independent of celestial signatures, e.g. integrating data per detector over many different pointings and time for sufficient statistics (typically one orbit). 
Re-determining this detector ratio in successive time intervals, e.g. per each orbit can accommodate long-term changes.
\item[(b)] 
Intensity variations of instrumental background components among adjacent pointings can be traced through an integrated signal. This is obtained from all detectors and over a wider energy range, and thus has sufficient statistical precision. The statistical limitations of data from only a single detector and a fine energy bin can be overcome in this way. 
\end{itemize}

\noindent
Therefore we decompose background into the physically-different model parts of \emph{spectral signature, detector ratio signature,} and \emph{background intensity variation} per each component $n$:
\begin{equation}
 B_{n,i,j}= t_i \cdot r_j \cdot s_{j,n} 
\label{eq:bgd_factorisation}
\end{equation}
{Here, $i$ characterises time in units of pointings, and $j$ identifies individual detectors.}

The \emph{spectral signature s} for background lines essentially is a Gaussian with a centroid energy that is characteristic for the physical process. This needs to be adapted to each detector's spectral resolution and degradation, and their temporal variations (see Equ.~\ref{eq:line-function} below for more details). These detector properties can be determined from combining information of many instrumental lines. The individual-line results can be combined with constraints for their consistent behaviour with energy, which will reduce any systematic variations that could otherwise be introduced from statistical uncertainty of a local measurement per detector and energy bin. 

The \emph{detector ratio signature r} characterises background types that originate from physical processes linked to specific isotopes, and their locations within spacecraft and instrument (see Equ.~\ref{eq:det-ratio} below for more details). It is is determined from all data integrated over the energy range corresponding to the line width of the particular process, accounting for detector-specific spectral responses.    
The solid angle under which a particular SPI detector 'sees' the relevant nuclei, isotopes, and $\beta$-decays affects its count rate for a particular background process. 
The effective solid angle\footnote{Here, \emph{effective solid angle} means taking into account both the emissivity of the source and the shielding effect of intervening material.} of each detector for such background is given by geometry of detectors,  instrument, and spacecraft; all this remains constant. 

Changes in the primary irradiation of the materials by cosmic-ray particles will affect the overall intensity of each type of background event, but not the locations of their origins. The cosmic rays as originators of the background penetrate all materials, and thus the characteristic detector ratio will remain constant, although the total intensity may vary (see Sect. 5 below). 
Second-order effects might be due to the preferred incidence directions of cosmic rays, especially at lower energies where rigidity cutoffs are dominant, e.g. when the satellite orientation is changed  by major angles, pointing to a different sky region. 
But, most importantly, the relative detector pattern does not change from pointing to pointing (see Figures~\ref{fig:SPI-detratio-lines} and \ref{fig:SPI-detratio-times}), in particular not among the different dithering pointings when a particular sky region is observed for $\sim$days.
The particular pattern of relative counts among detectors for instrumental background thus can be determined in better statistical precision from data accumulating over multiple pointings. 

The \emph{intensity variation} $t$ over the times of consecutive pointings can then be determined, eliminating the known effects of the different detector responses and relative detector intensity ratios. Integrating over data from all contributors with analogous characteristic intensity variations between different pointings, statistical precision is increased for this relative change of background. 
This is done fitting for each successive pointing the total intensity of a spectral model of background consisting of all lines and continuum, summed over all detectors with their specific responses. 
Note that the total (absolute) number of background counts per spectrum finally needs to be determined according to 
 Equ.~\ref{eq:model-fit} above, also accounting for sky contributions. 

This factorisation of background (Equ.~\ref{eq:bgd_factorisation}) allows to determine each of its components on a broader data basis, compared to background modelling that relies on each single energy bin's counts, with their statistical (Poissonian) uncertainty. 
Thus the overall precision of the background model is improved. 
The statistical uncertainty, however, in each data space bin, remains a fundamental limit (see Equ.~\ref{eq:model-fit}). 

\section{Determining spectral-response and background characteristics and trends}  
\label{sec:mission-analysis}

We exploit the first 13.5 years of the INTEGRAL mission data archive towards establishing a detailed spectral response and background history of SPI: The instrumental background dominates SPI data and includes a multitude of clear lines, which hold the necessary details. 
We use Ge detector data in the energy range 20-2000 keV. These are binned at a resolution of 0.5 keV, which is small compared to the instrumental resolution of SPI of, e.g., { 2~keV (FWHM) around 0.5 MeV}. 
With respect to temporal resolution, a compromise is required between best spectral precision through sufficient statistics, and choices of accumulation times for avoidance of any systematic variations or distortions.
Spectral shapes and intensity of background vary over time, and the spectral detector responses change from degradation and annealings;  all this blurs spectral signatures within a few keV in data integrated over longer times.
Fig.~\ref{fig:SPI-spectrum_all-mission} shows the all-mission spectrum (above) and sub-samples for one orbit and one detector only (below). 

As an initialisation step, we identify all spectral features above a smooth continuum using the cumulative all-mission spectrum (see Fig.~\ref{fig:SPI-spectrum_all-mission}, top). 
This accumulation maximises statistical precision so that even weak lines can be identified, although lines may be somewhat distorted in shapes. 

\begin{figure}
  \centering
  \includegraphics[width=\linewidth]{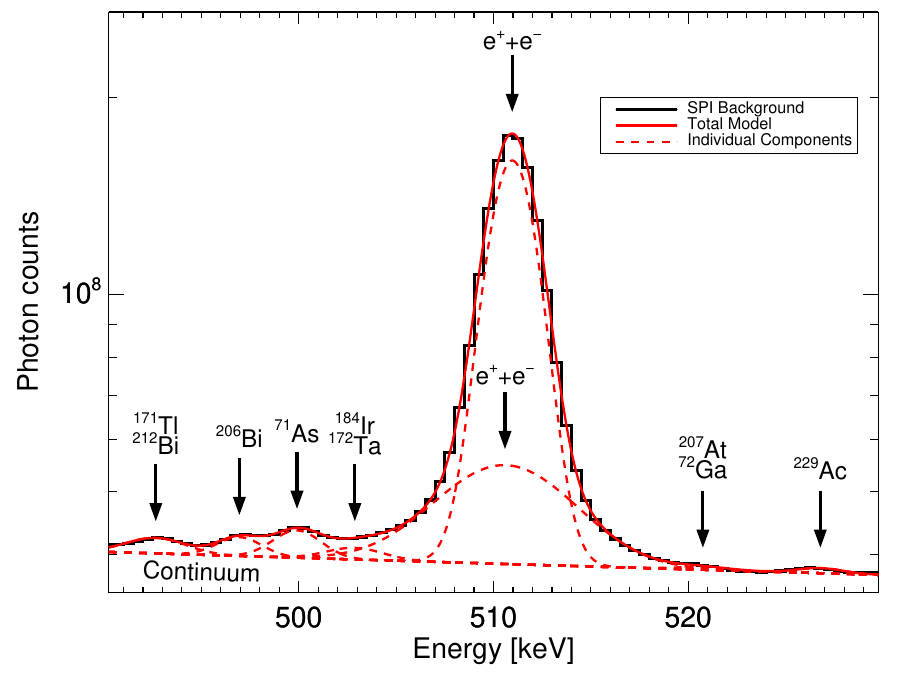}
    \includegraphics[width=\linewidth]{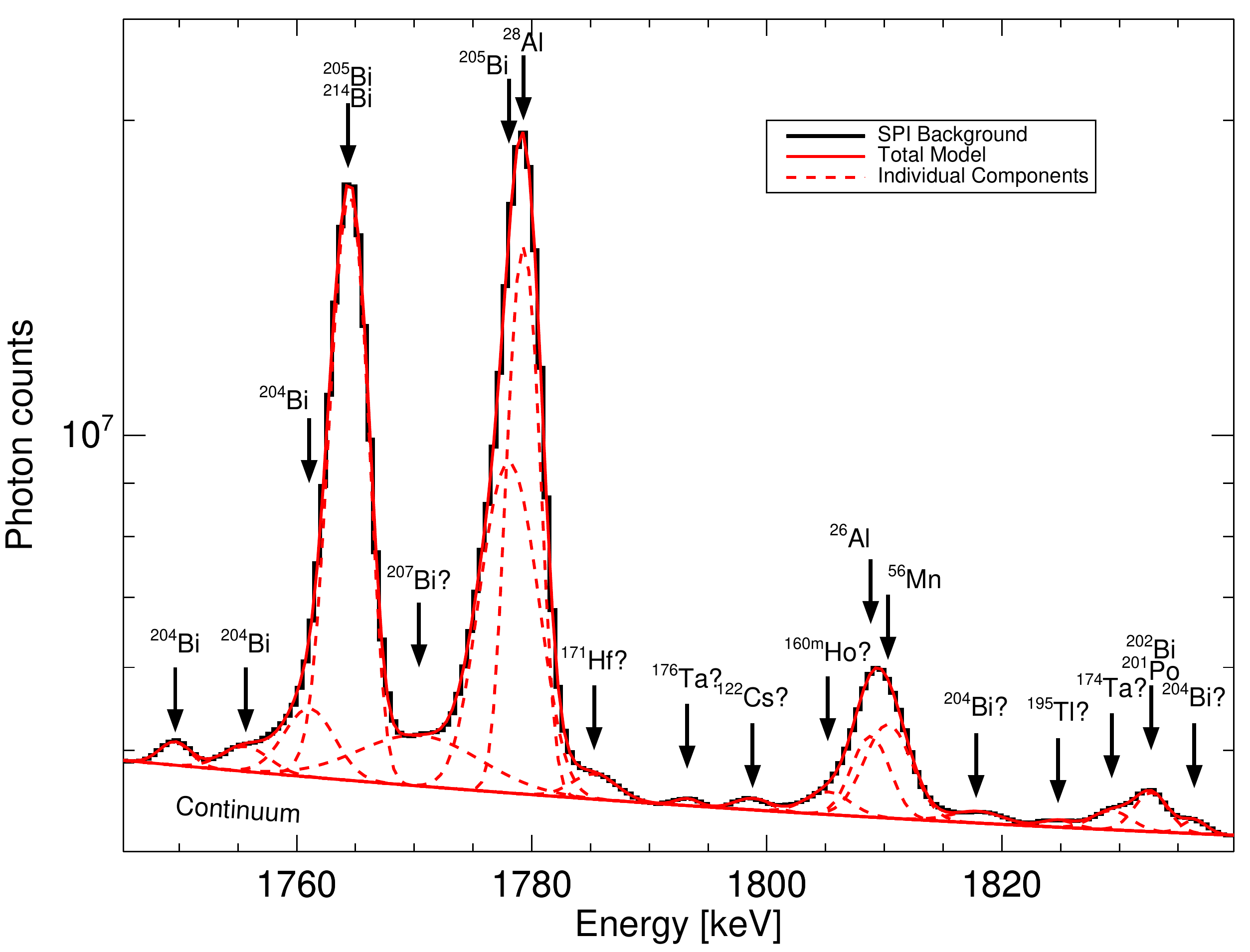}
   \caption{{ Spectral decomposition into lines,} near the positron annihilation line at 511 keV (\emph{above}), and the  $^{26}$Al line (\emph{below}). The annihilation line itself requires a composite model of a narrow (instrumental width) and a broader component, representing different modes of $\beta$-decays in spacecraft or detector materials. Several low-intensity lines at nearby energies are part of the spectral model, all modelled as Gaussians \emph{(red dashed lines)}. The isotopes causing these various background lines, as much as identified, are indicated.}
  \label{fig:SPI-spectrum_all-mission_511_1809}
\end{figure}


Each line is now modelled as a convolution of a Gaussian line profile $G(E)$ { with amplitude $A_0$, centred at energy $E_0$ and with a width parameter $\sigma$}, with a one-sided exponential $T(E)$, the latter representing the degradation of charge collection with time between annealings { as charcterised with degradation parameter $\tau$} \citep{Kretschmer:2011,Siegert:2017}:
\begin{eqnarray}
G(E;E_0,\sigma) & = & A_{0} \exp \left(- \frac{(E-E_{0})^2}{2 \sigma^2} \right) \label{eq:gaussian} \\
T(E;\tau) & = & \frac{1}{\tau} \exp \left( \frac{E}{\tau} \right) \quad \forall E > 0 \label{eq:tail} \\
L(E;E_0,\sigma,\tau) & = & (G \otimes T)(E) = \nonumber \\
& = & \sqrt{\frac{\pi}{2}} \frac{A_{0} \sigma}{\tau} \exp \left( \frac{2 \tau (E-E_{0}) + \sigma^2}{2 \tau^2} \right) \nonumber \\
& & erfc \left( \frac{\tau (E-E_{0}) + \sigma^2}{\sqrt{2} \sigma \tau} \right)
\label{eq:line-function}
\end{eqnarray}

We represent broader spectral features with sets of such one-side-distorted Gaussians. Therefore, not all candidate lines in our list are related to real physical processes, at this stage.  
{ In particular, Compton edges of strong lines may thus show up as, or blend with, other lines in our list\footnote{We flag such possible Compton edge origin in column 4 of the line table in the Appendix.}.}   
The underlying continuum is modeled as a power-law distribution, normalised at the centre energy $E_m$ within the region of interest:
\begin{equation}
\centering
 C(E;\alpha, c_0)  =  c_0 \left( \frac{E}{E_m}\right)^\alpha 
 \label{eq:continuum}
 \end{equation}
 The complete list of identified lines is presented in the Appendix, Table~\ref{table_SPI-line-table}.

Fig.~\ref{fig:SPI-spectrum_all-mission_511_1809} shows in detail how specific parts of the spectrum can be represented in this way. The spectral model over a range of $\sim$35--140~keV is composed of a minimum number of lines of instrumental width. These line models are connected, as their shape parameters share the detector characteristics of resolution and degradation. 
Thus, e.g. for a 120-keV wide region, now a model with, e.g., 20 lines has only parameters of amplitude and centroid per each line, plus spectral resolution, degradation, and two continuum parameters, i.e. $20*2+2+2=46$ parameters; for comparison, modelling by the original fine spectral bins in this case would result in $240$ background-related parameters.

These spectral model components, as defined from the full-mission integrated spectrum, are fit now to a dataset accumulated during a shorter time. { Here, a compromise must be chosen between statistical precision and tracing temporal variations. The spectrum should have lines which should be free of distortions from degradation or calibration drifts, so that it represents the spectral response properly. Then, each line should have sufficient statistics so that Poissonian statistical uncertainties are small compared to the systematic variations from trends such as degradation or increase from cumulative activation. 
With a degradation of about 1\% per orbit and a spectral resolution of about 1--2\%, an accumulation time of one orbit is our preferred choice, and compromise. Towards higher energies beyond 1.4~MeV and their lower count rates, we found necessary to integrate spectra from groups of several orbits, however.}

We fit a region of interest (typically a range of about 100 keV, with typically 20-30 components), separately for each individual detector. This results in spectral-line and spectral-response parameters for each detector and orbit, including line centroids ($E_0$) and amplitudes, and spectral shapes ($\sigma, \tau$) which characterise its spectral response.

\begin{figure}
  \centering
  \includegraphics[width=\linewidth]{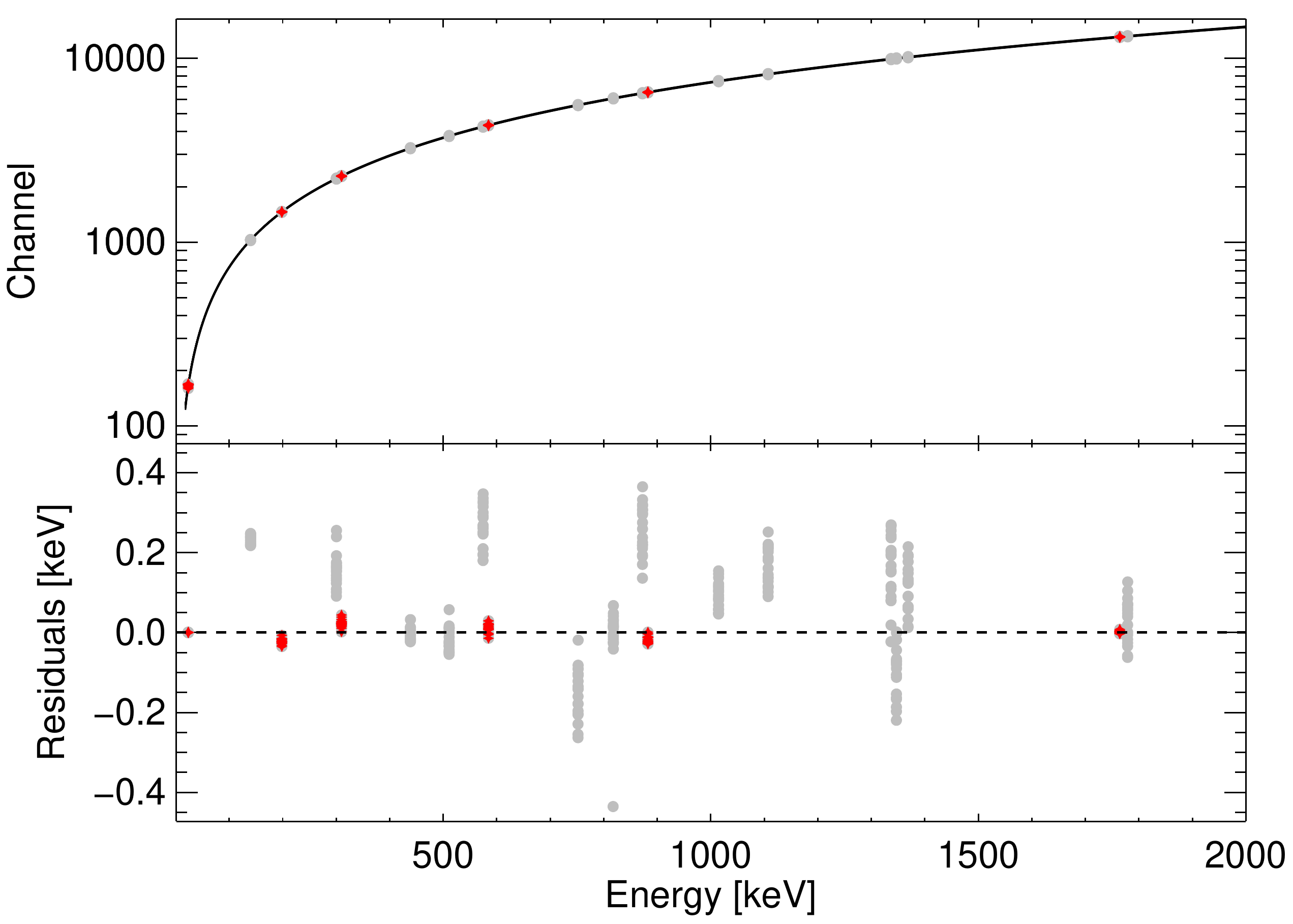}
   \caption{Energy calibration of line centroids, channel units versus energy. The energy calibration function used in preprocessing is shown as (\emph{solid line in upper graph, dashed line in lower graph}), and is derived from Gaussian fits to a few strong lines which are marked in \emph{red}. Data points show the centroids as extracted from our fits, for the strongest single/isolated background lines.  The lower graph shows the deviations of each line centroid from the expected energy.}
  \label{fig:energyCalibration}
\end{figure}
 
\begin{figure}
  \centering
  \includegraphics[width=\linewidth]{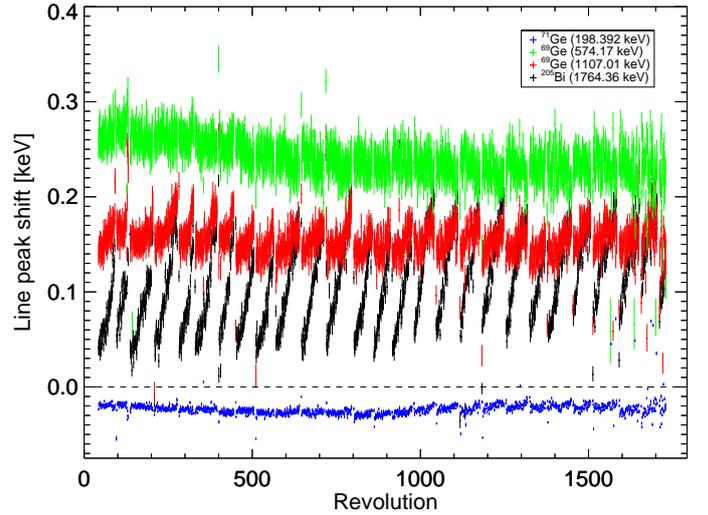}
  \caption{Variations of line centroid energies for four strong lines { with time. Each data point is the result of a fit for data from one orbit.} Each line apparently has its own systematic offset, superimposed onto variations of 0.01 to 0.03~keV. Degradation effects are evident, strongest for the $^{205}$Bi line. { The corrections of orbit-by-orbit gain variations in preprocessing, by using centroids of fitted symmetric Gaussians, leads to gradual shifts of calibration peaks, which then is observed as positive energy shift in the asymmetric line fits reported here.}}
  \label{fig:energyCalibrationStability}
\end{figure}

We validate the spectral fit quality through its $\chi^2$ value, and the resulting detailed spectral parameters undergo a consistency analysis to remove outliers and glitches from degeneracies. For example, resolution widths $\sigma$ and degradations $\tau$ together determine the line widths, and present a degeneracy for data with higher statistical noise. 
Neighbouring lines and other detectors allow consistency checks, and the 
recognition of outliers leads us to iterations of the fit with constrained parameters, to evaluate such degeneracies and resolve them towards a smooth and consistent trend of spectral response parameters.
Also, at this step any transient sources that might have significant contributions above instrumental background, such as solar flares or the flare from microquasar V404 Cygni, will be recognised, and corresponding data will not be used for determination of the instrumental response.
 The results of these spectral fits for the entire mission are maintained in a database, which we provide as a service to interested scientists. We describe the database file contents and access tools to extract parameter values of interest in the Appendix below.

\begin{figure*}
  \centering
  \includegraphics[width=0.8\linewidth]{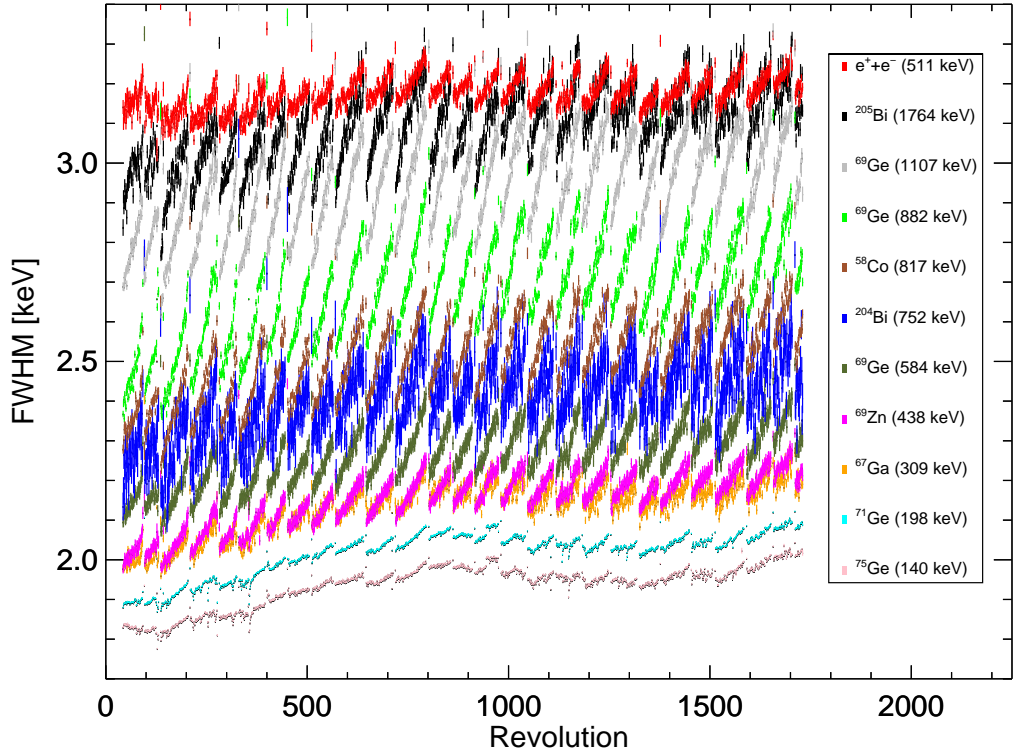}
   \caption{{ Time} evolution of the spectral resolution. Shown are the line widths for several instrumental lines (see legend), with energies between 198 and 1764 keV, as fitted per data from one orbit ('revolution').}
  \label{fig:spectralResolution}
\end{figure*}
 

\section{Spectral-response properties} 
\label{sec:spec-response}

The  energy calibration from electronic pulse height units to keV units is derived from line centroids of Gaussian-shape fits for six strong lines  between 100 and 1800 keV. Gain variations from orbit to orbit also are accounted for,  based on the observed shifts of their centroids between orbits; the calibration reference is taken from the early part of the mission, from orbit 43. This energy and gain calibration is applied in pre-processing, and used throughout all later data analysis.
Fig.~\ref{fig:energyCalibration} shows how the centroid fits to the strongest, single, isolated background lines with our multi-component spectral model Equ.~\ref{eq:line-function} compares to this calibration of energies.
The calibration function of energy $E$ [keV] versus pulse height $p$ [channels 1-16384] is modelled as
\begin{equation}
E =c_1 \frac{1}{p}+c_2 + c_3 p +c_4 p^2
\label{eq:Ecalib}
\end{equation}
(see solid line in Fig.~\ref{fig:energyCalibration} upper graph).
The absolute energies associated to the reference line centroids are taken from the associated physical processes, i.e. line origins assigned to the most plausible isotope \citep{Firestone:1998,Weidenspointner:2003}.
The six strong calibration lines are marked (red color) in Fig.~\ref{fig:energyCalibration}; see also Table~\ref{table_SPI-line-table}). 

Degradation of charge collection produces an asymmetric line shape, and the spectral peak appears at lower energies with increasing degradation. This determines the energy calibration used in pre-processing, and so our spectral fits recover these distortions, as an apparent increase of fitted photo peak centroid energies, when this degradation asymmetry is accounted for, Equ.~\ref{eq:line-function}. 
The selection of six lines only for determination of a four-parameter calibration curve also leads to some scatter, that can distort the calibration curve over a wider energy range (see Fig.~\ref{fig:energyCalibration}). 
It is evident that each line has its specific systematic offsets, from line blends or other physical effects that may shift the line centroid energy away from the tabulated literature value, or from variations of the gain correction. These offsets are within a few tenths of a keV. 
The intrinsic (statistical) precision of the SPI energy calibration is about 0.02~keV, as estimated from the variations of derived centroids of the 198~keV line. The achieved absolute precision depends on the particular line, and is rather estimated as $\pm$0.04 to 0.05~keV, when accounting for systematic variations incurred by the inadequate treatment of degradation in the calibration and gain correction during pre-processing
 (see Fig.~\ref{fig:energyCalibrationStability}).    
  
We recover the proper absolute energy knowledge for each individual $\gamma$-ray line in our spectral fits, as we account for detector resolutions and degradation. We thus obtain more realistic line centroids, which can be combined with physics knowledge of specific line energies for proper absolute calibration of energy in our spectra. These should be used when absolute energy units are important. 
For example, a Doppler shift of 0.1 keV in the { astrophysical} $^{26}$Al line corresponds to a kinematic velocity of 122 km~s$^{-1}$ \citep{Kretschmer:2013}. 
The effective line broadening from these effects in data integrated over many revolutions is a second-order effect and modest, by comparison with the statistical uncertainty of fitted line widths. 
 
\begin{figure}
  \centering
  \includegraphics[width=\linewidth]{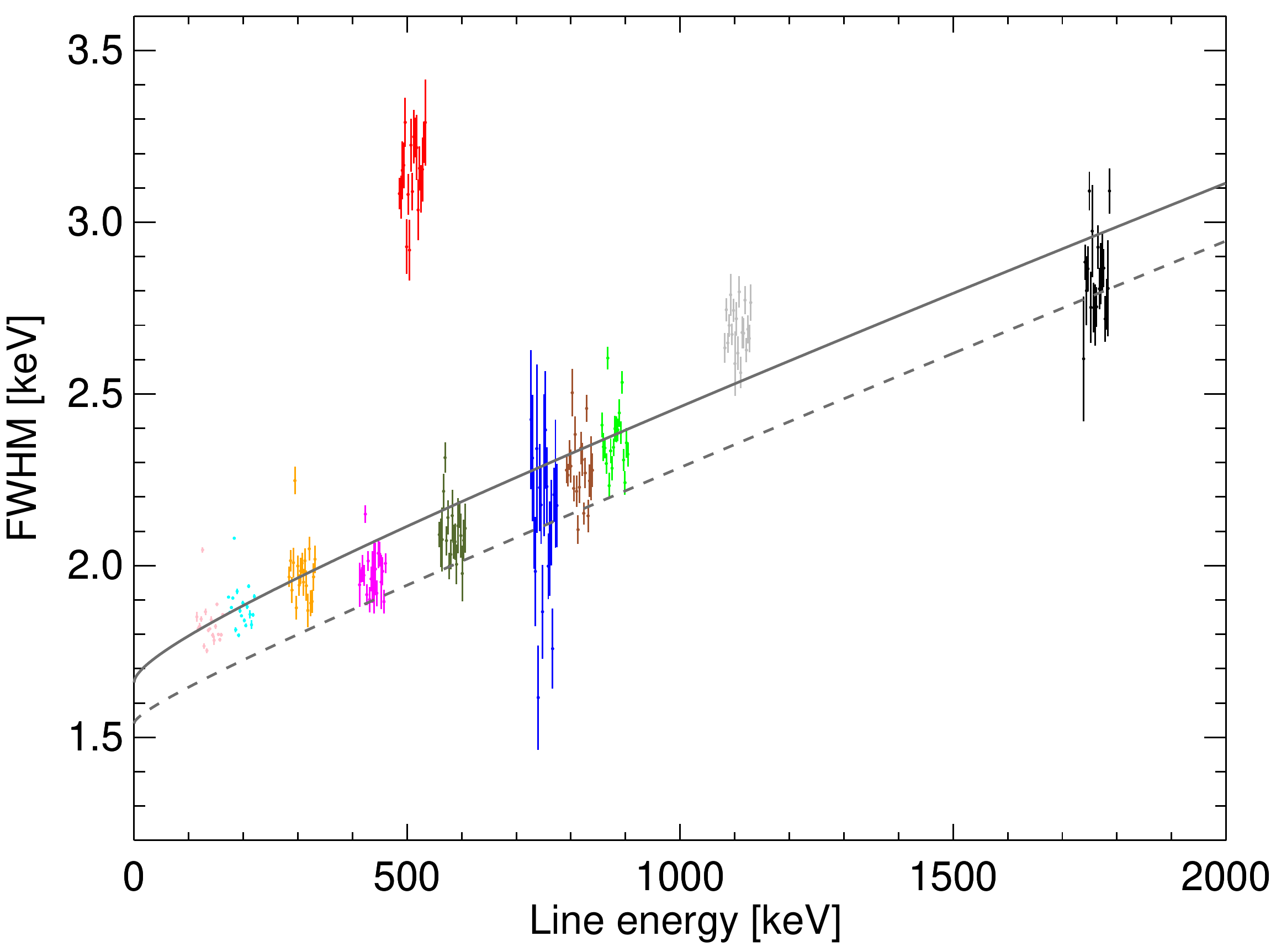}
   \caption{{ Energy} resolution versus energy. { Shown are the line widths (FWHM) for} the strongest single/isolated background lines. The energy resolution functions are shown as obtained from pre-launch calibrations \citep{Attie:2003} (\emph{dashed line}), and from mission data  (\emph{solid line}). Results obtained for different detectors are slightly offset in energy for better visibility, and are centred around the laboratory energy of the respective background line. The 511~keV line clearly presents an anomaly, { and the 1107 keV line from $^{69}$Ge may also be blended with contributions from nearby lines} (see text).}
  \label{fig:fwhm_all-lines}
\end{figure}

\begin{figure}
  \centering
  \includegraphics[width=\linewidth]{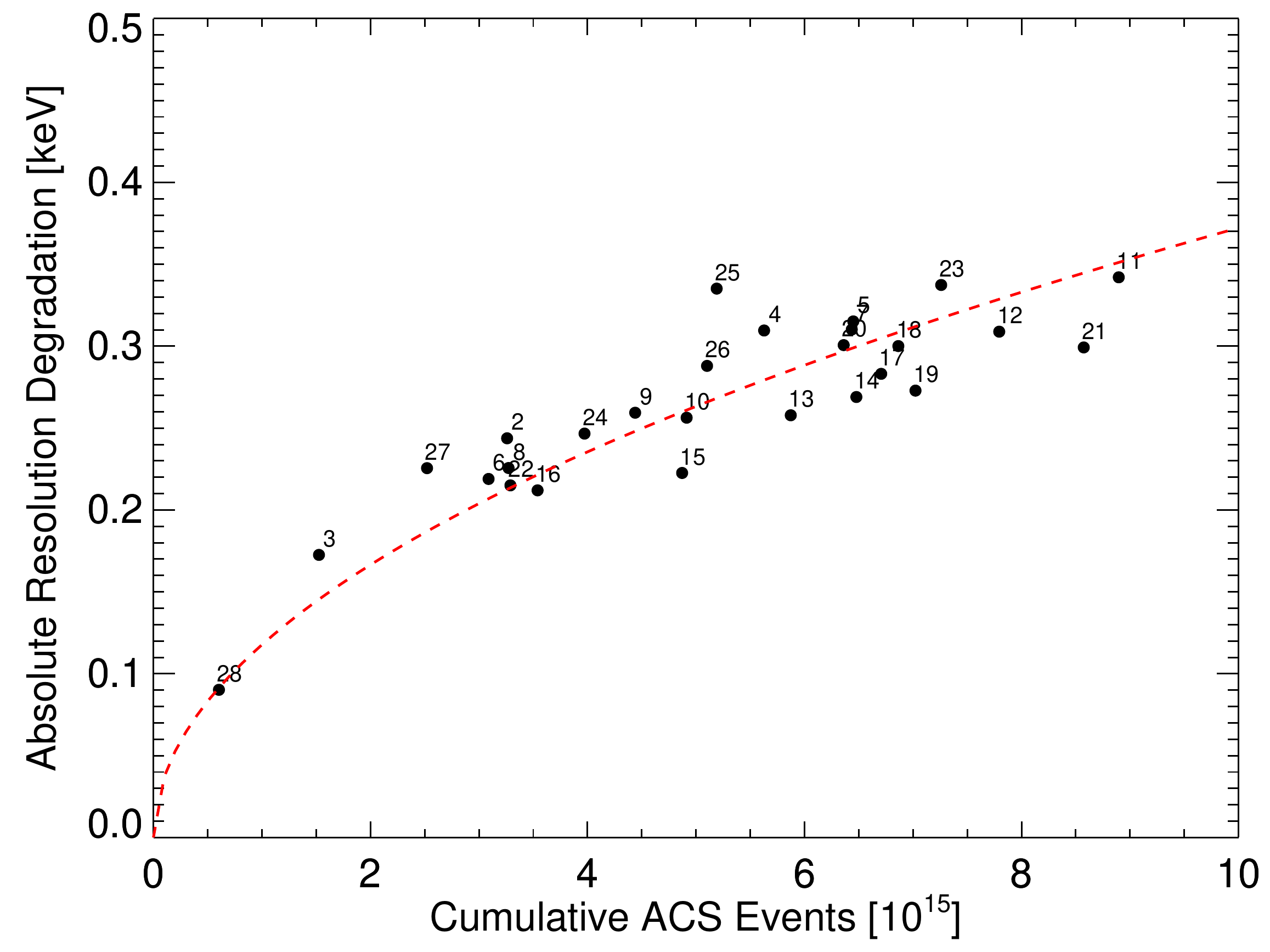}
   \caption{{ Degradation of spectral resolution versus integrated charged-particle dose.} The integrated veto detector system counts between previous and current annealing cycle { are used as a measure of the charged-particle dose, as the} veto detector system mainly counts charged-particle interactions, which should scale with the incident cosmic-ray flux. Data points are labelled with the sequential number of the annealing cycles. A relationship $y= (0.118\pm0.002)\sqrt{x}$ fits degradation  responding to charged-particle bombardment (\emph{dashed line}).}
  \label{fig:degrad-activation}
\end{figure}

\begin{figure}
  \centering
  \includegraphics[width=\linewidth]{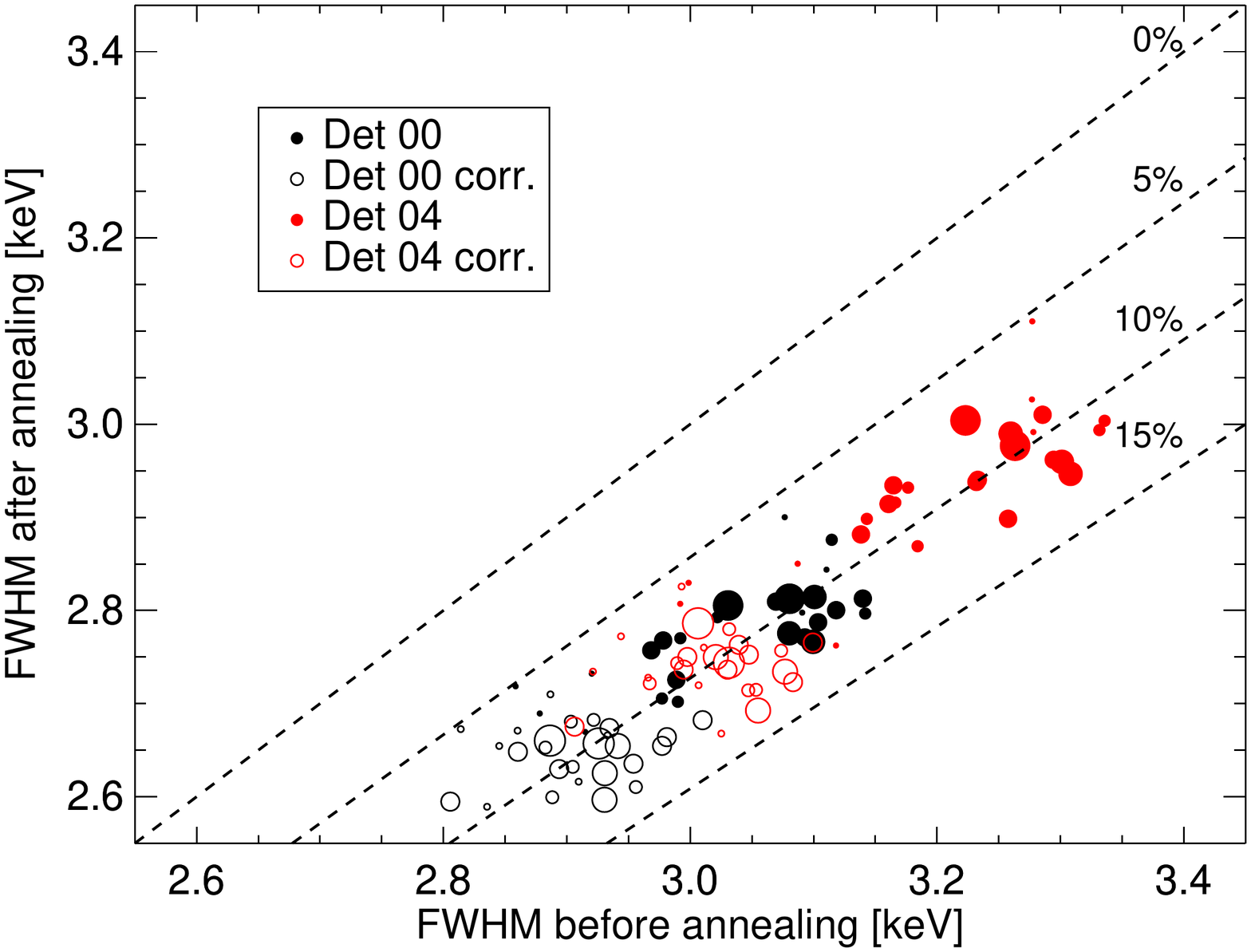}
   \caption{{ Recovery} of spectral resolution for each annealing cycle. The abscissa shows the level of degradation at the onset of the annealing operation, the ordinate shows the resolution as achieved immediately after the annealing, each for one of the better and one of the poorer resolution detectors (0,4). \emph{Filled symbols} are data points using the FWHM of the 1107~keV line before and after each annealing; \emph{open symbols} are the same data, corrected by eliminating a long-term trend. The size of symbols represents the duration of anealings (see text).}
  \label{fig:recovery-dets}
\end{figure}

Fig.~\ref{fig:spectralResolution} shows the evolution of spectral line width (FWHM) over 13.5 years, for several strong lines  between 100 and 1800 keV. 
The spectral resolution degrades about linearly between annealings { by 10--15\%}. The degradation rate between different annealings generally varies with the overall background intensity as reflected in the anticoincidence system rates (Fig.~\ref{fig:SPI-bgd-rates}), and shows some irregular variations, 
generally remaining within a narrow range of 0.5--1.5\% per revolution. Spectral resolution is restored from its gradual degradation with the annealings in a single step. 
This behaviour is seen in all lines and at all energies. Absolute degradation increases with increasing line energy (see Fig.~\ref{fig:spectralResolution}).

We only use lines for analysis of the spectral response which  are \emph{stand-alone} and \emph{single} lines, i.e., unaffected by blends, and thus approximate most closely the intrinsic detector response function
see Table~\ref{table_SPI-line-table}, where these \emph{response-calibration lines} are highlighted).
It may not always be possible to clearly isolate instrumental response behaviour, 
and our visual inspection for  \emph{stand-alone} lines may not always ensure that we are free of contamination in our sample of intrinsic response measurements. 

The spectral resolution of SPI's Ge detectors across the energy range up to 2 MeV is shown in Fig.~\ref{fig:fwhm_all-lines}. 
The expected trend with energy due to the increasing number of charges liberated with increasing energy deposit in the Ge crystal is shown for comparison, both as obtained from the prelaunch calibration (dashed line; \citet{Attie:2003}), as well as fitted to the mission data (solid line). 
Deviations from an idealised model of a Gaussian spectral response result in systematic deviations from the overall energy trend for specific lines. 
 It is evident that the 511~keV line is systematically broader; it includes components of physically-different origins (see the two components identified in Fig.~\ref{fig:SPI-spectrum_all-mission_511_1809}).
Also, the detectors with intrinsically poorer resolution stand out (e.g., det. no.~4).



The SPI team pioneered the regular annealing operations for maintaining high spectral resolution of a Ge detector in space \citep[see][and references therein]{Lonjou:2005,Fahmy:2008}. 
Annealing operations are initiated when degradation is significant but still moderate, because the exercise relies on thermal reversibility of the loss in charge collection.
Figures~\ref{fig:degrad-activation} and \ref{fig:recovery-dets} show how annealings succeeded in recovering the resolution from its gradual degradations, over the 13.5 years of the mission. 
The level of degradation should scale with the cumulative cosmic-ray bombardment. In Fig.~\ref{fig:degrad-activation} we therefore show the absolute increase in resolution (FWHM) of the 1107~keV background line versus the cumulative count rate of the SPI anticoincidence system, which measures the charged-particle flux incident on SPI in orbit. A square-root increase of degradation with charged-particle dose (dashed line) is evident, { and would be expected if 'exposure' dominates the effect}.

In Fig.~\ref{fig:recovery-dets} we compare resolution before and after the annealing, for two representative detectors. In this graph, the size of data points per annealing represents the duration of the annealing; values were between 36 hours early in the mission and 200 and up to 225 hours for most of the later annealings. We also show resolution improvements per annealing in a second set of data points where we eliminated the apparent long-term trend (see Fig.~\ref{fig:spectralResolution}). For this we fitted as a $\sqrt(x)$ relation through resolutions as obtained after each annealing, and extract only the steps in resolution between two successive annealings (open symbols). 
We conclude that all anealings achieve a comparable and satisfactory recovery of spectral resolution. Differences between detectors are larger than the effects of total degradation suffered before the annealing was initiated. Also  impacts of annealing duration are rather insignificant; estimates prior to the mission were that a duration of two days would already fully recover a 20\% loss of spectral resolution \citep{Leleux:2003}.

%
\section{Instrumental-background properties}
\label{sec:background}

The general long-term variation of instrumental-background intensity is illustrated in Fig.~\ref{fig:SPI-bgd-rates} for detector count rates. 
Our detailed spectral fits extract more-specific  information about different  background components.  
We will now identify background lines with common underlying physics.  Such lines should share properties such as intensity variations, and thus can broaden the data base and help to overcome statistical limitations.  
We explore temporal variations of intensities and characteristic detector ratios, combining this with correlations among different lines across the energy spectrum.

\begin{figure}
  \centering
  \includegraphics[width=0.9\linewidth]{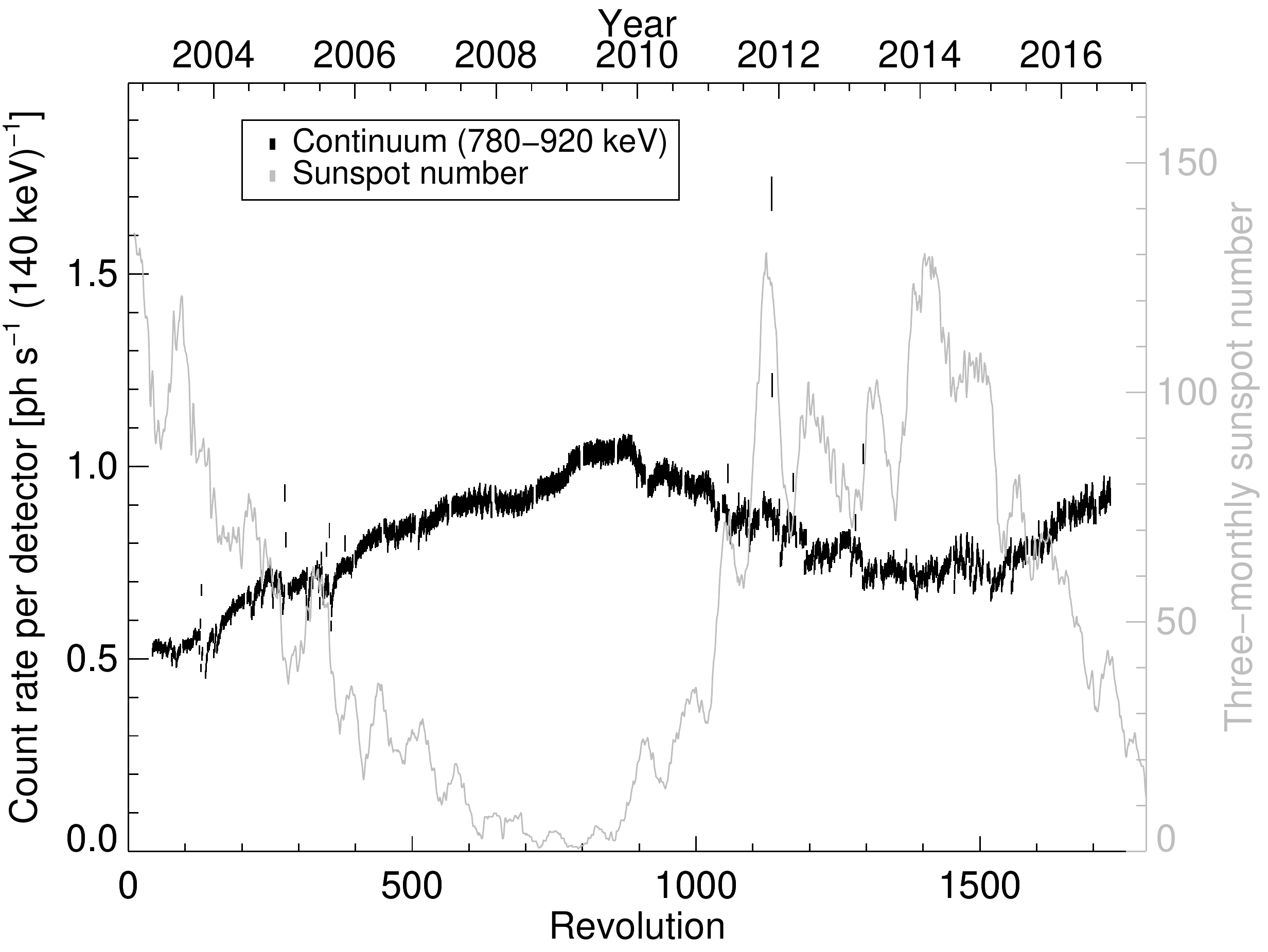}
    \includegraphics[width=0.9\linewidth]{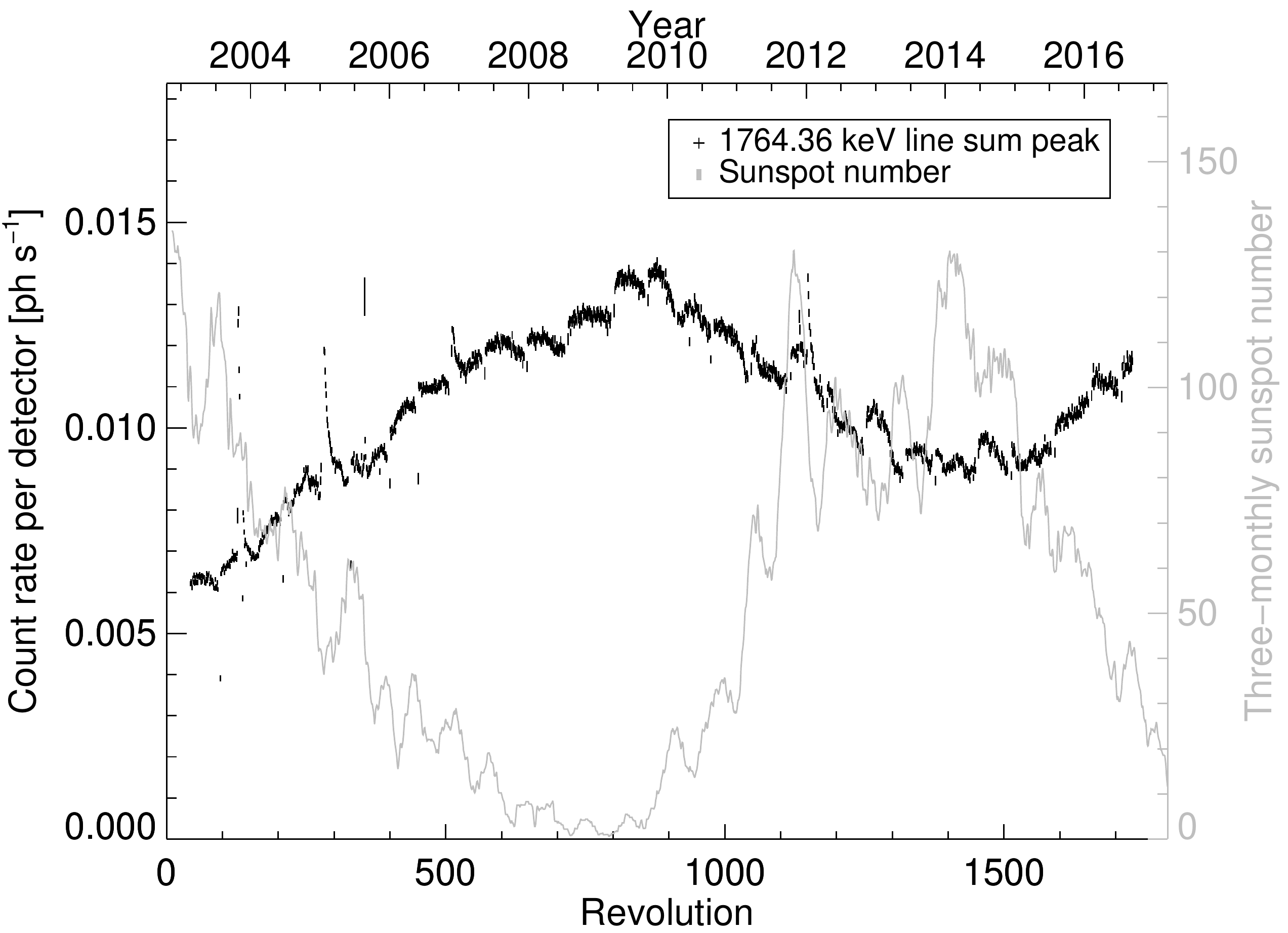}
      \includegraphics[width=0.9\linewidth]{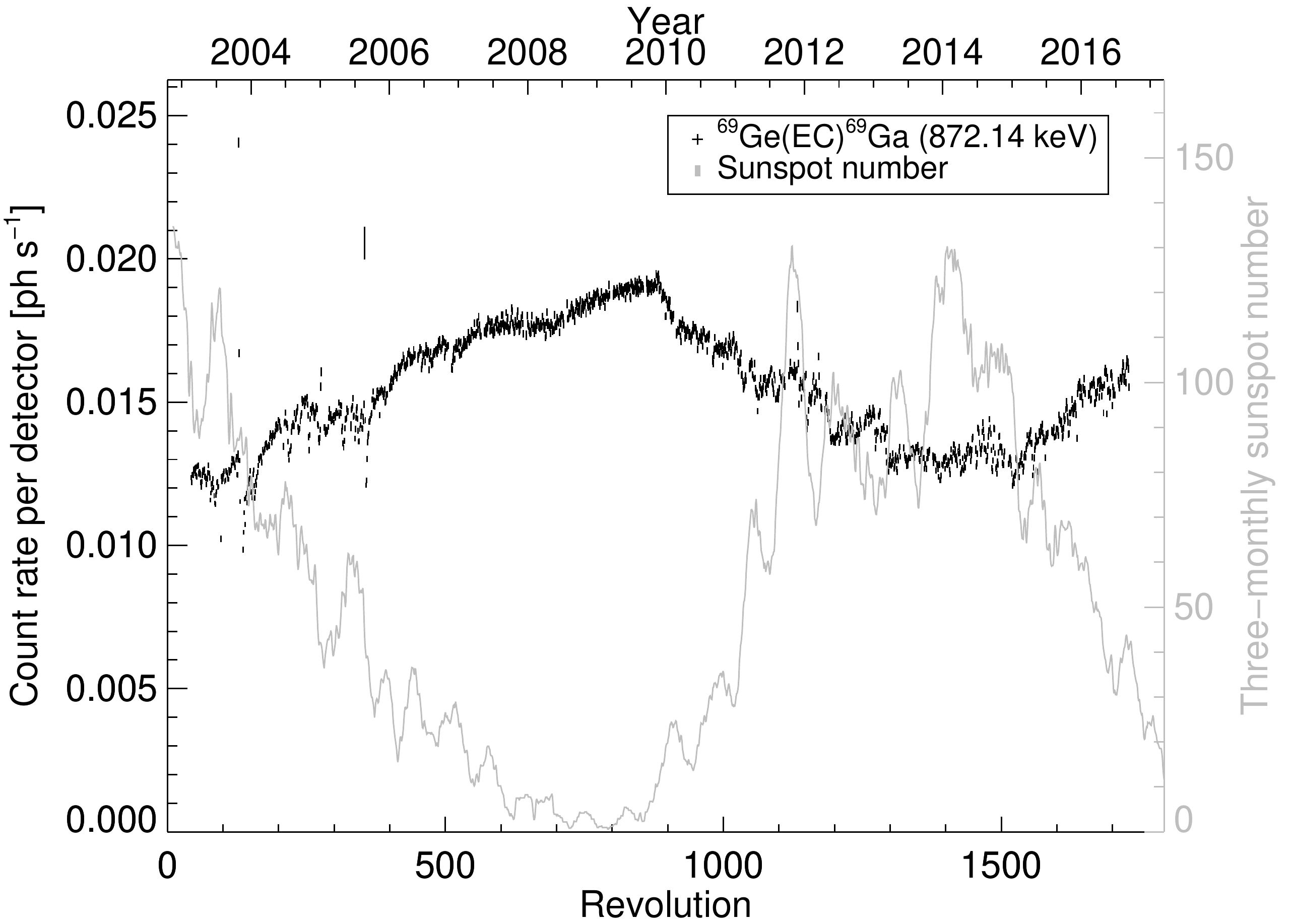}
   \caption{ Time evolution of background component intensities.  {\it Top:}  The underlying smooth continuum (here in a band 780-920~keV). {\it Middle:} $^{206}$Bi activation of anticoincidence material, through the 1764~keV line. {\it Bottom:} $^{69}$Ge activation within the detectors, through the 872~keV line. }
  \label{fig:bgdvariations_rates}  
   \end{figure}

\begin{figure*}
  \centering
    \includegraphics[width=0.4\linewidth]{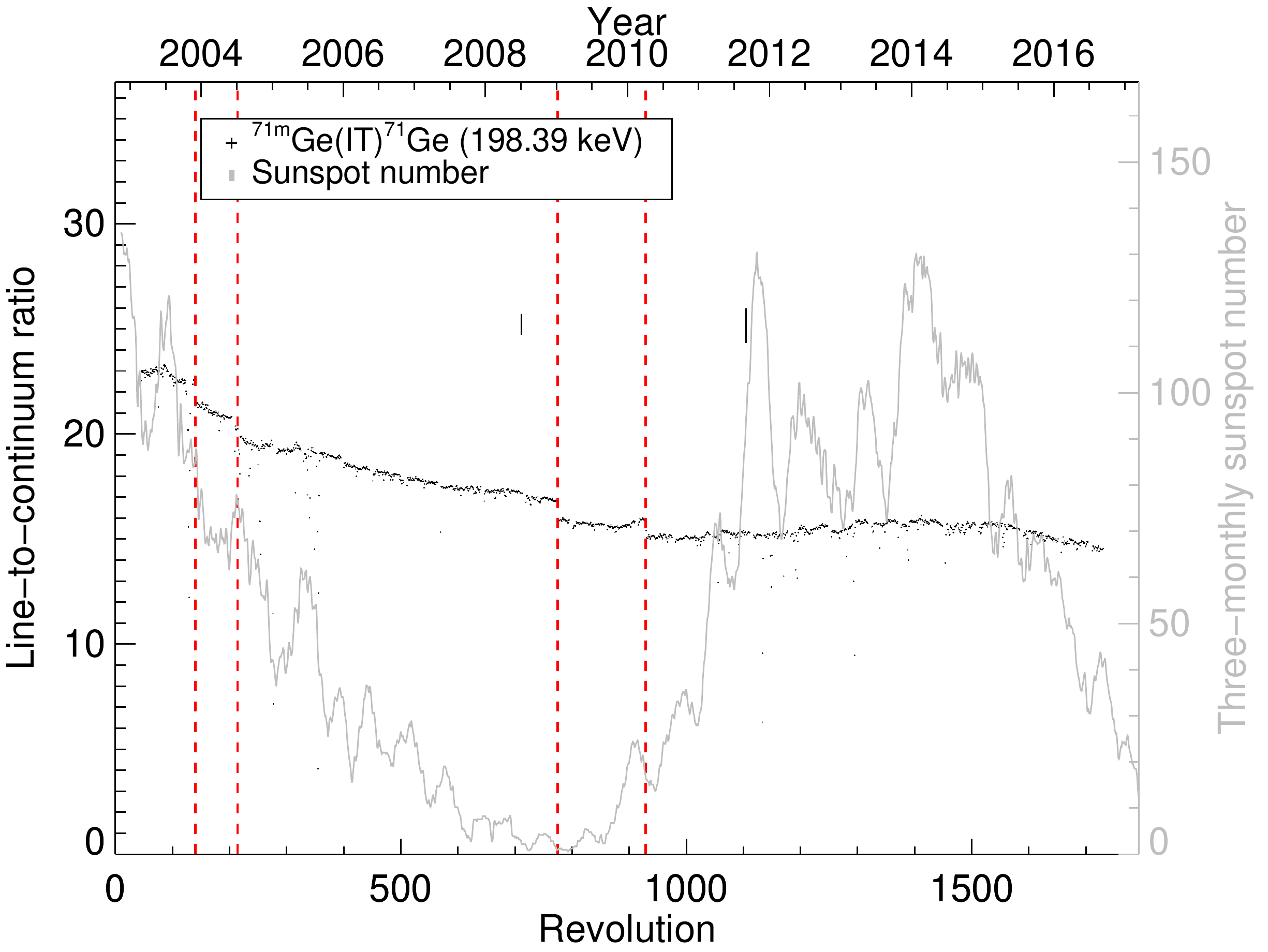}
        \includegraphics[width=0.4\linewidth]{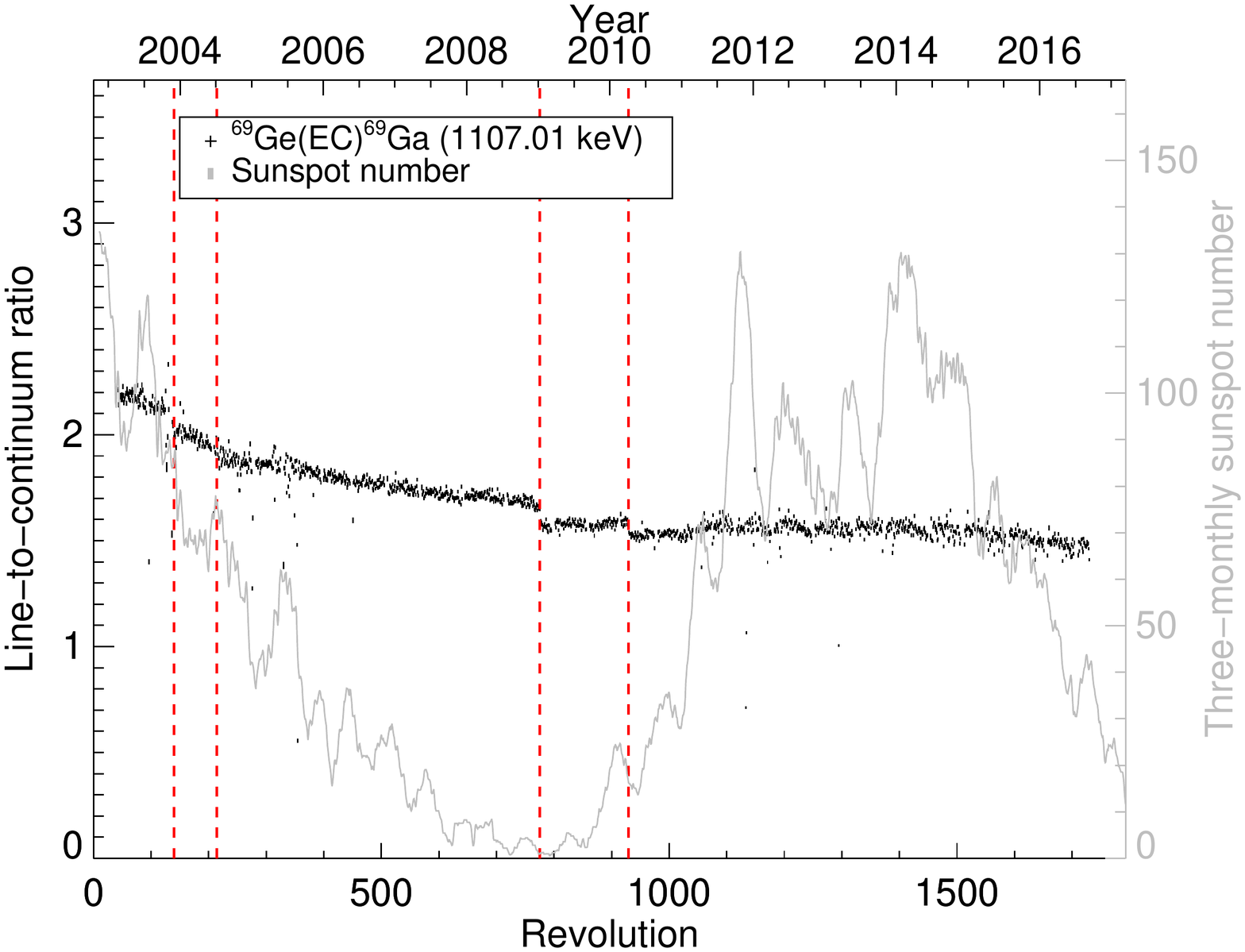}
     \includegraphics[width=0.4\linewidth]{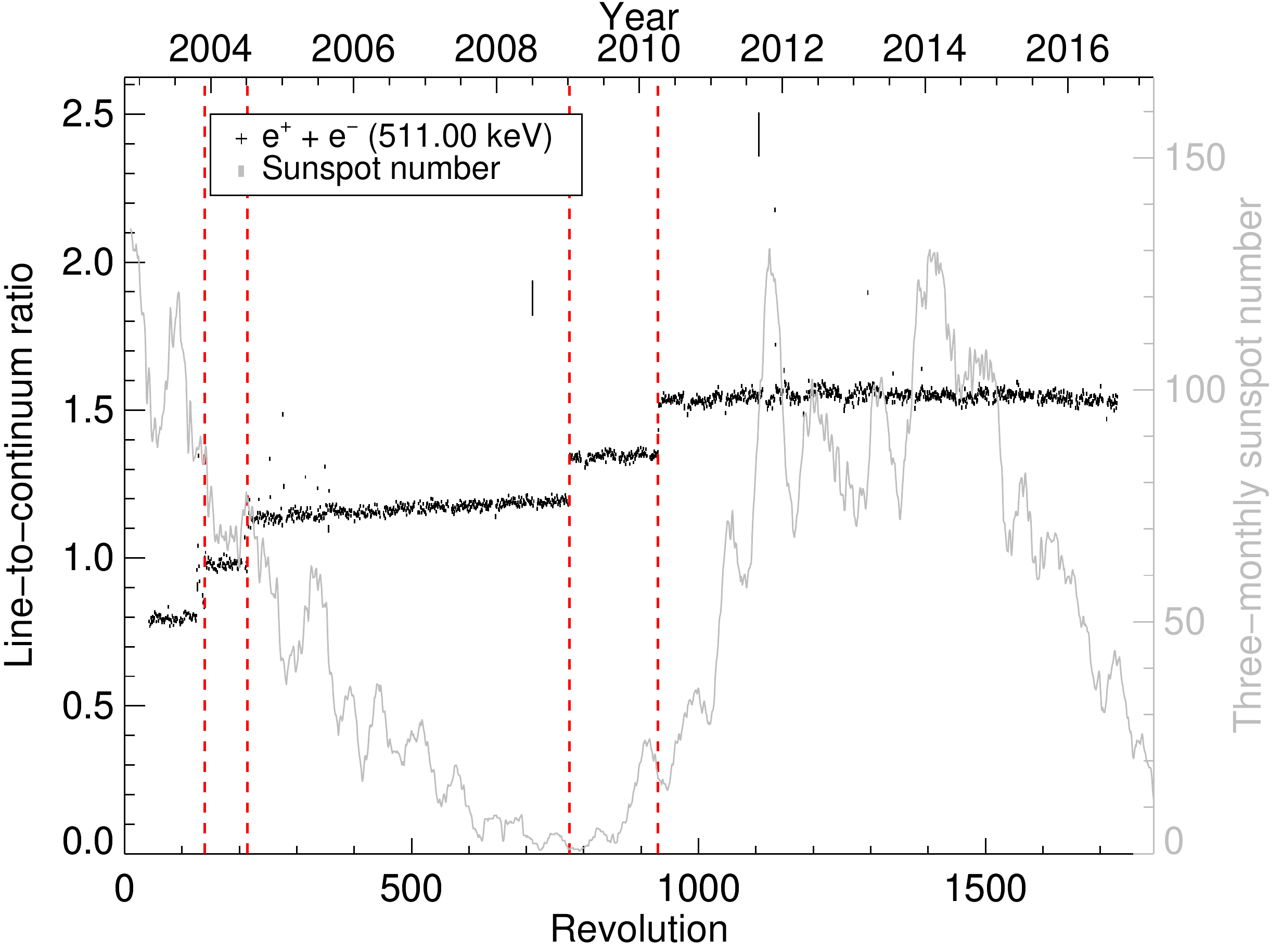}
           \includegraphics[width=0.4\linewidth]{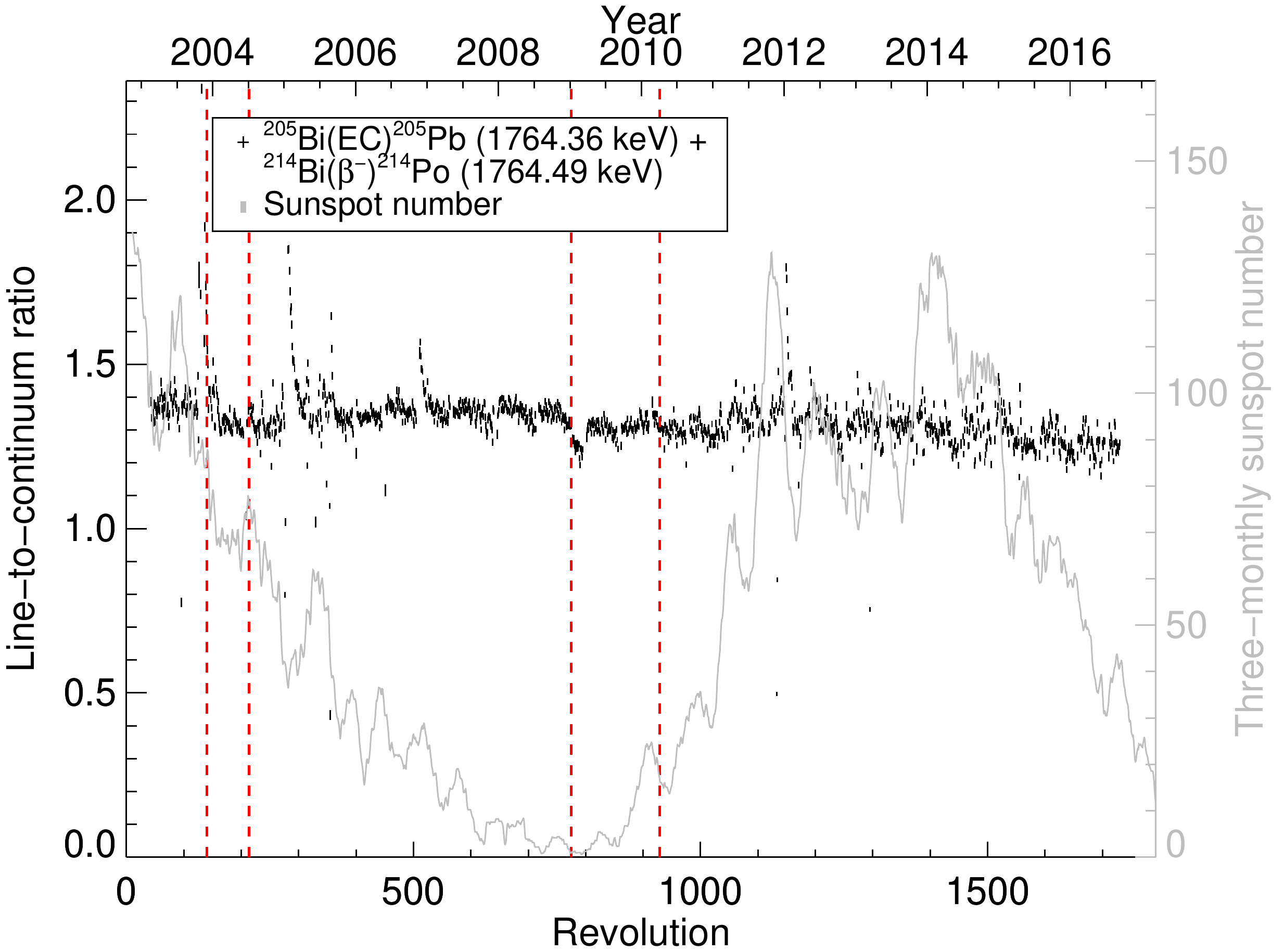}
      \includegraphics[width=0.4\linewidth]{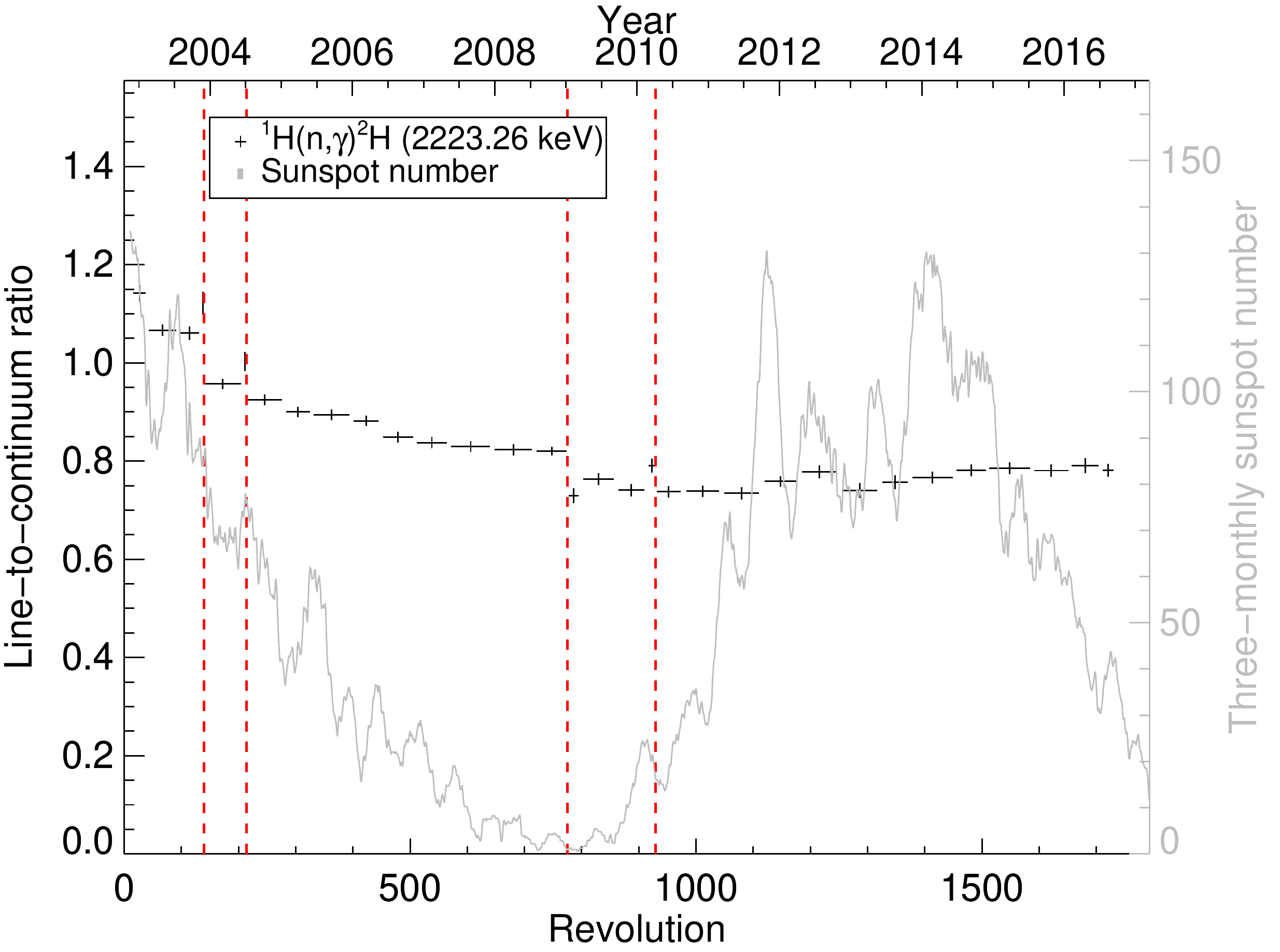}
       \includegraphics[width=0.4\linewidth]{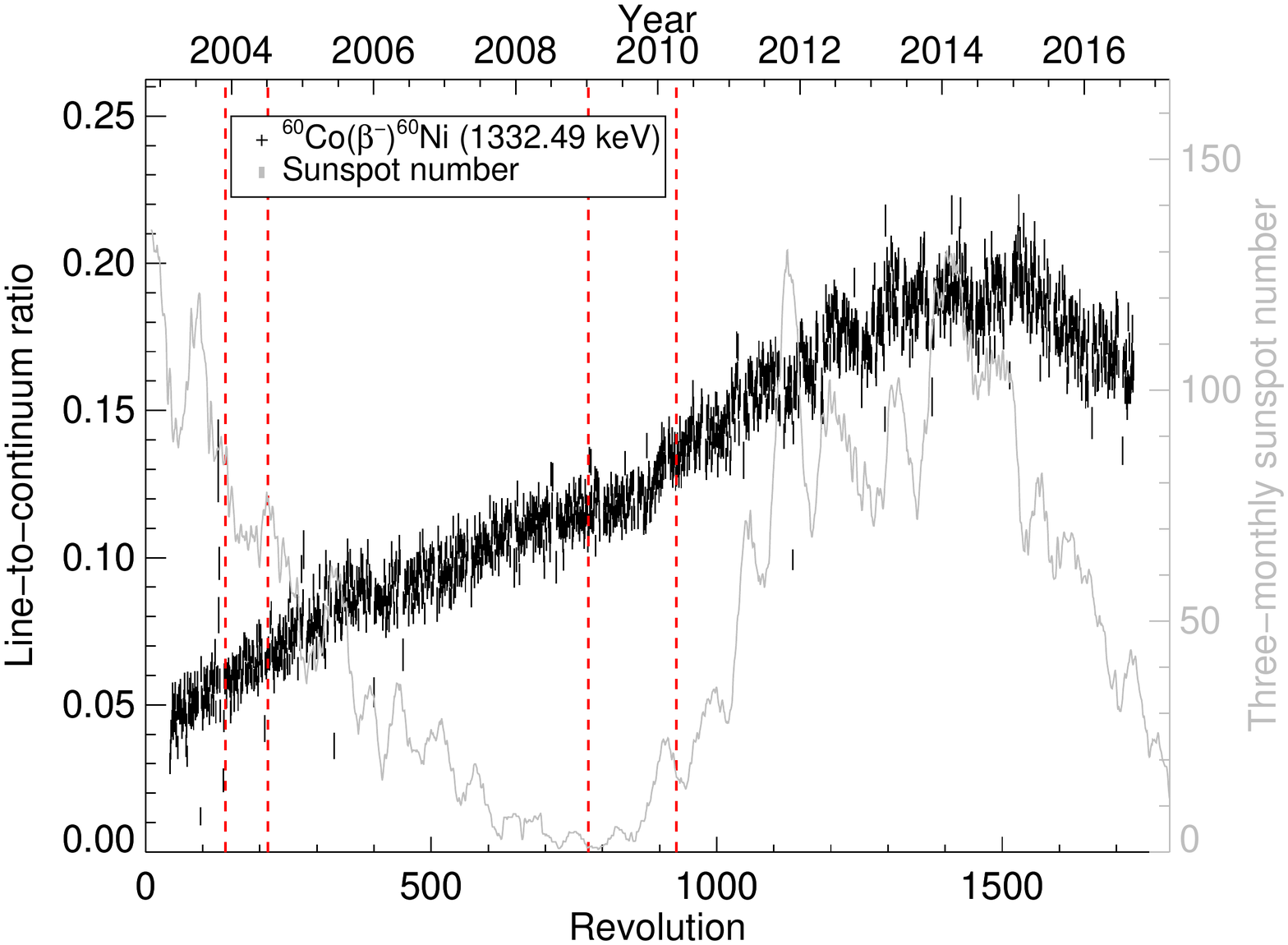}
    \caption{Temporal evolution of the intensities for different background lines. {\it Left column, from top to bottom:} The evolution of  the 198 keV line measuring neutron irradiation, the 511 keV line from local positron annihilation, and the neutron capture line at 2.2~MeV. {\it Right column, from top to bottom:} The 1107 keV line from $^{69}$Ge activation,  the 1764 keV line from $^{214}$Bi in the veto system, and the 1332 keV line from $^{60}$Co radioactivity as it builds up.  Intensities are scaled to underlying continuum intensity. For comparison, the time history of the sunspot number count is shown. Failures of individual Ge detectors occurred at times marked by dotted vertical lines, and show the most-significant impact for the case of the 511~keV line (see text).}
 \label{fig:bgdvariations_LineContRatios}
\end{figure*}

\subsection{Background intensity variations}

Fig.~\ref{fig:bgdvariations_rates} shows the mission history for the intensity of the continuum under the instrumental lines in the energy band 780--920~keV, and for two characteristic instrumental background lines. 
The variations seen in detector count rates (Fig.~\ref{fig:SPI-bgd-rates}) are again evident, with a maximum of intensities in the vicinity of revolutions 800--900, and a decline after that. In Fig.~\ref{fig:bgdvariations_rates} we also overplot the sunspot number count history (from http://sidc.be/silso/datafiles), which is a proxy for solar activity. 
The evident anti-correlation plausibly shows that cosmic ray intensity is reduced as the more-active Sun extends the heliosphere and increases magnetic shielding in INTEGRAL's orbit \citep{Potgieter:2013}. 

Gamma-ray lines have specific origins, while the continuum includes a mix of different background physics. 
This explains differences in detail in the variations of different background components.
In the examples shown in Fig.~\ref{fig:bgdvariations_rates}, the smooth continuum and background lines originating in Ge and Bi materials, respectively, show different amplitude dynamics and detailed reactions to solar flares.  
 
In Fig.~\ref{fig:bgdvariations_LineContRatios}, line intensities are scaled to the intensity of the fitted continuum under the line itself (a band of 3 times the line width $\sigma$, centred on the line energy), to extract the differences between continuum and a specific line. Shown are the characteristic cases of the 198 keV line measuring neutron irradiation, the 511 keV line from local positron annihilation, the neutron capture line at 2.2~MeV, the 1107 keV line from Ge activation,  the 1764 keV line from Bi in the veto system, and the 1332 keV line from build-up of $^{60}$Co.  
The time history of the sunspot number count, shown also in these graphs, reflects solar activity. Apparently the correlation seen above largely disappears through the normalisation with continuum intensity.  
Specific behaviour of individual lines becomes more apparent. 
The failures of individual Ge detectors occurred at times marked by dotted vertical lines, and coincide with step changes of different magnitude.
 The 511~keV line shows strongest effects from detector failures: 
 positron annihilation results in emission of two 511 keV photons in opposite directions; therefore, it is likely that a positron annihilation event triggers two neighbouring detectors, which sorts the triggered events into the class of \emph{multiple events}  (see Section 2.1 above). When a neighbouring detector is eliminated, therefore, the rate of \emph{single events} increases, as shown here. 
 The neutron activation background line at 198 keV (top left), as well as the neutron capture line (bottom left), both seem to show indications of a correlation with the solar activity cycle; as the continuum-intensity normalisation mainly characterises charged-particle background, this indicates that charged-particle and neutron background scale differently with solar activity.
For lines that result from charged-particle background, the characteristic variation (see Fig.~\ref{fig:bgdvariations_rates}) disappears, when scaled with underlying continuum. The same is true for the 511~keV line from positron annihilation, which appears constant in time when scaled with continuum intensity (middle graph in Fig.~\ref{fig:bgdvariations_rates}). 

\begin{figure}
  \centering
  \includegraphics[width=0.9\linewidth]{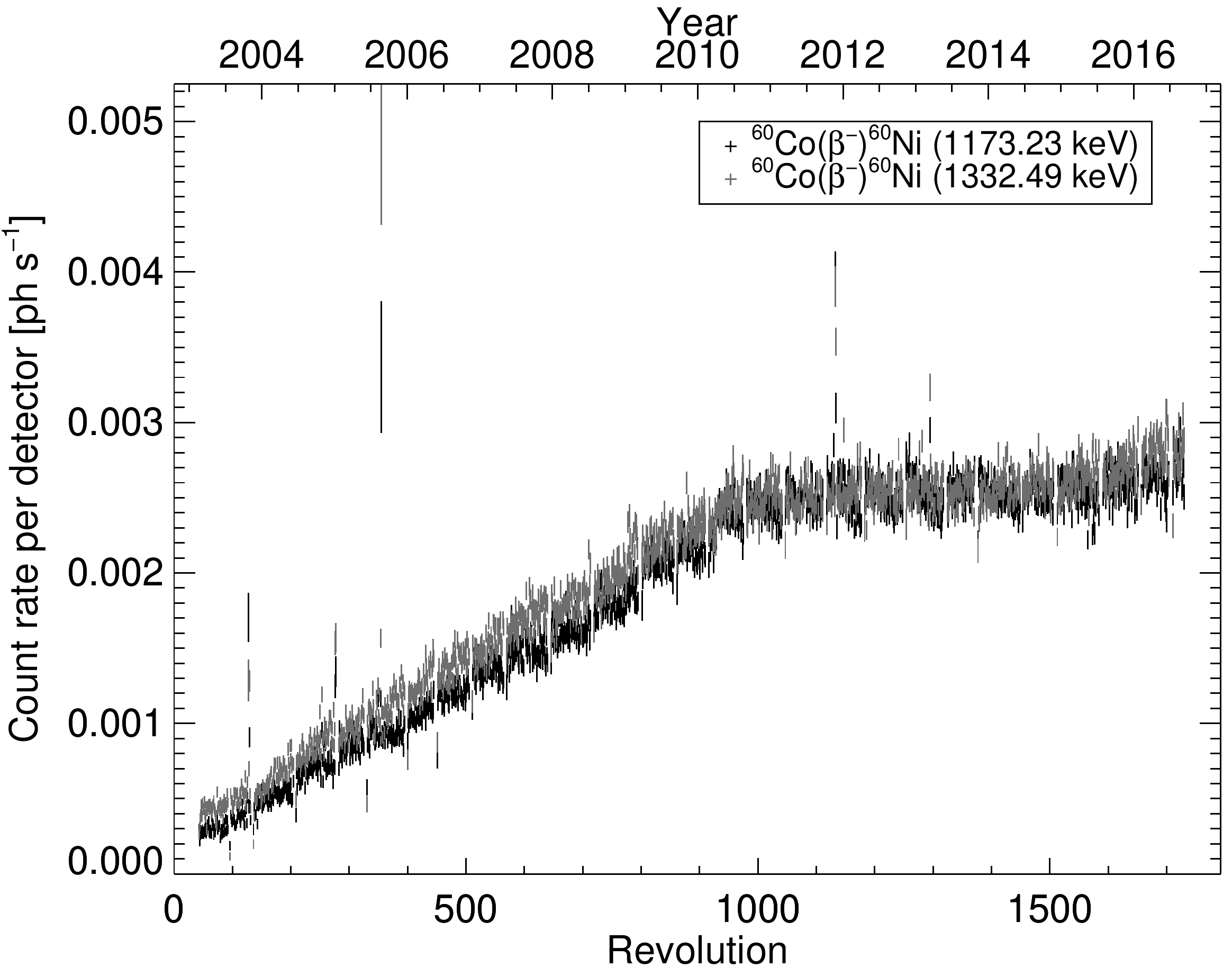}
        \includegraphics[width=0.9\linewidth]{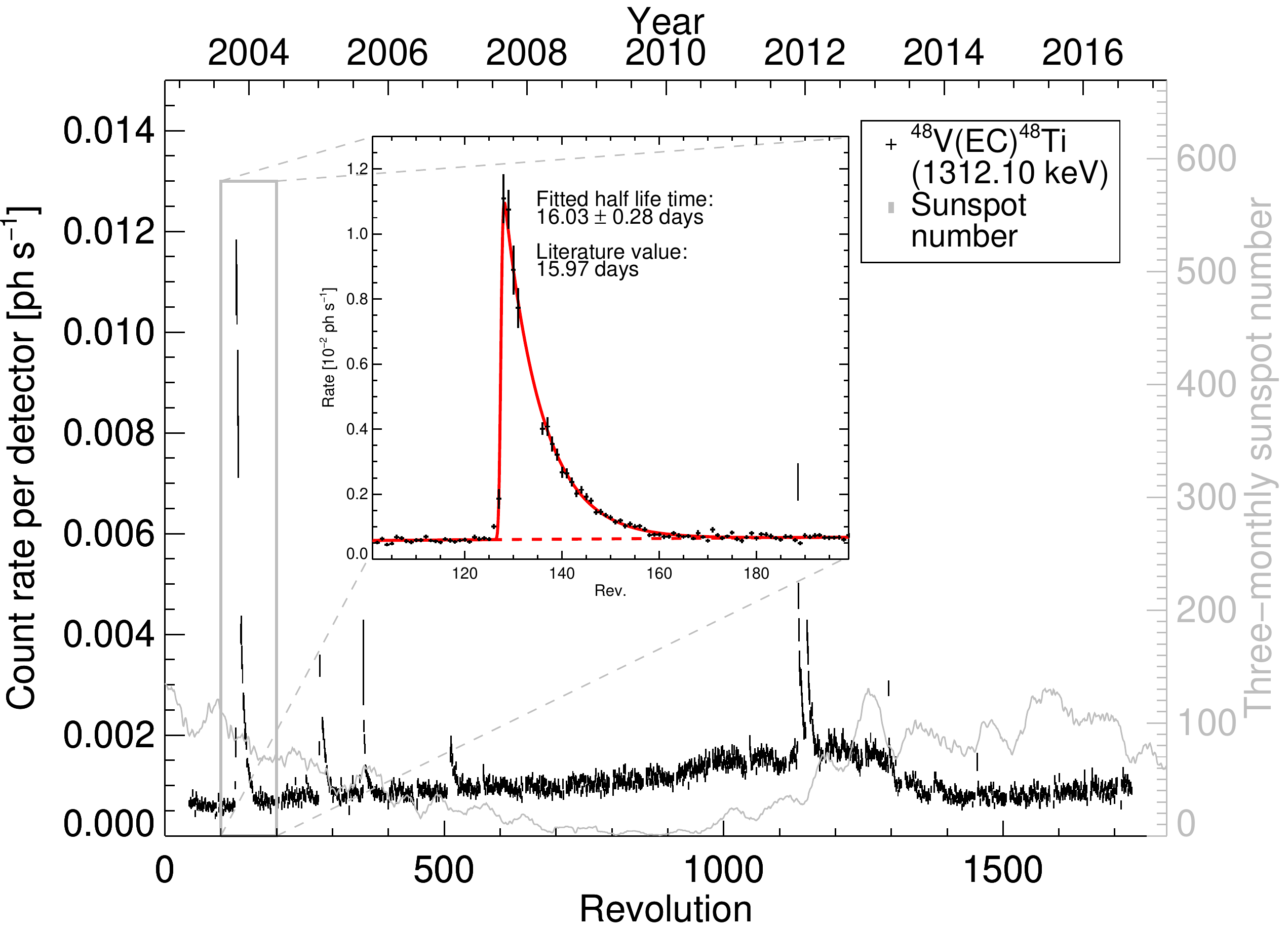}
   \caption{Time evolution of background rates for lines from radioactive species. {\it Top:} $^{60}$Co activation,  leading to gradual build-up due to the radioactive lifetime of 7.6 years. {\it Bottom:} $^{48}$V activation, with a decay time of 16 days. Exponential decay can be seen after an activation event (insert) { such as a particle storm following deformations of the Earth's magnetosphere from solar activity}; but the shorter decay time avoids longterm build-up such as seen in $^{60}$Co. }
  \label{fig:radioactbgd}
\end{figure}

Build-up of radioactivity with mission time is clearly seen for the $^{60}$Co line (bottom right);  this isotope with a radioactive lifetime of 7.61 years shows the characteristic long-term increase. 
$^{60}$Co decay leads to simultaneous emission of lines at 1173 and 1332 keV energy, in a cascade of de-excitation of the daughter nucleus. A closer look at the temporal evolution of these two lines is shown in Fig.~\ref{fig:radioactbgd} (upper panel): Indeed, both lines show a parallel build-up of intensity as  $^{60}$Co radioactivity is created in spacecraft material. 
The lower panel of Fig.~\ref{fig:radioactbgd} shows how an impulsive, short activation as it occurs when solar flare protons hit the spacecraft, in addition to the long-term trend of cosmic-ray irradiation discussed above. In such a particle flare sweeping across the spacecraft, the radioactive $^{48}$V isotope is produced in significant quantities. The decay follows the activation event of the bombardment with the characteristic radioactive decay time of 16 days.

\begin{figure}
  \centering
  \includegraphics[width=\linewidth]{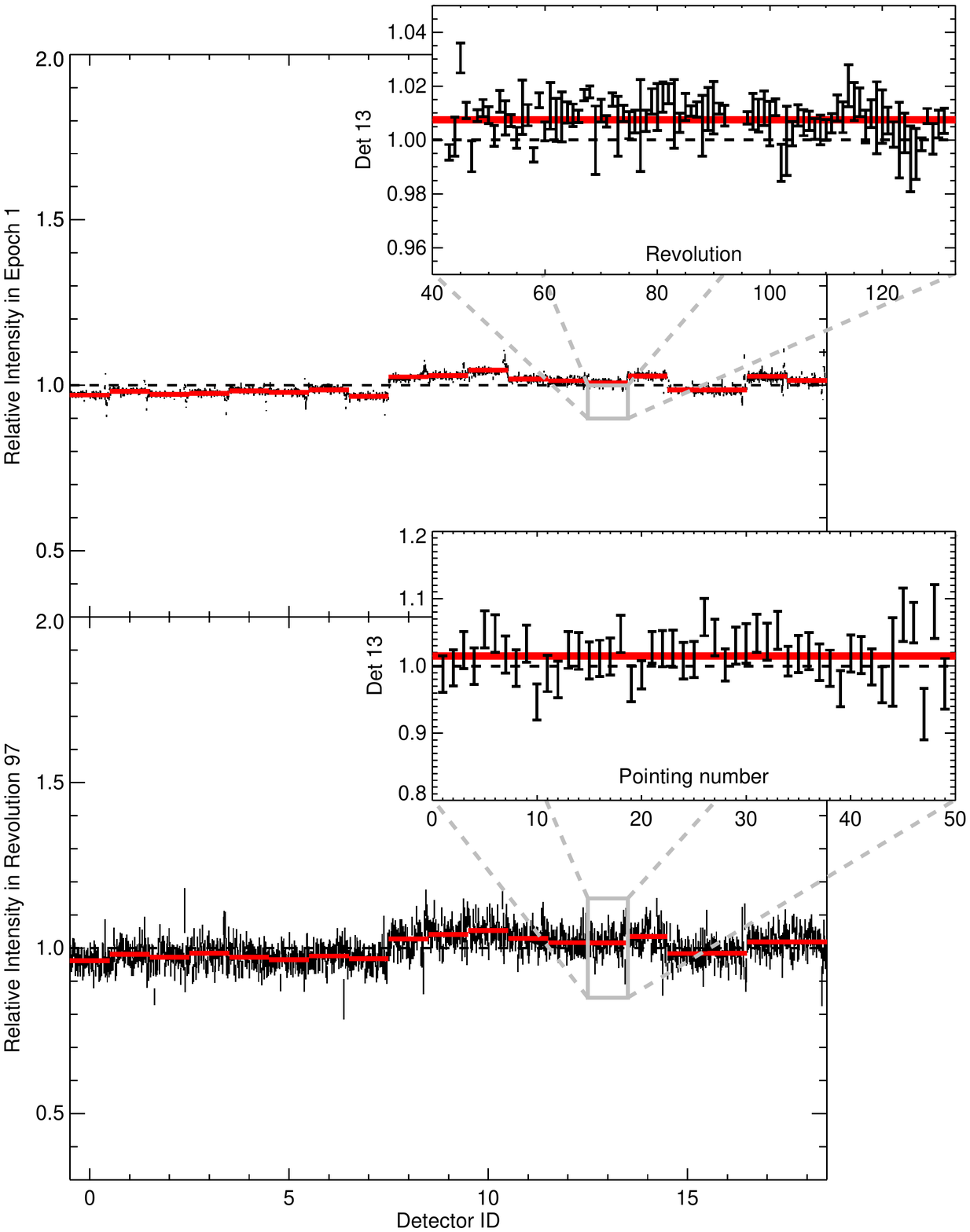}  
    \includegraphics[width=\linewidth]{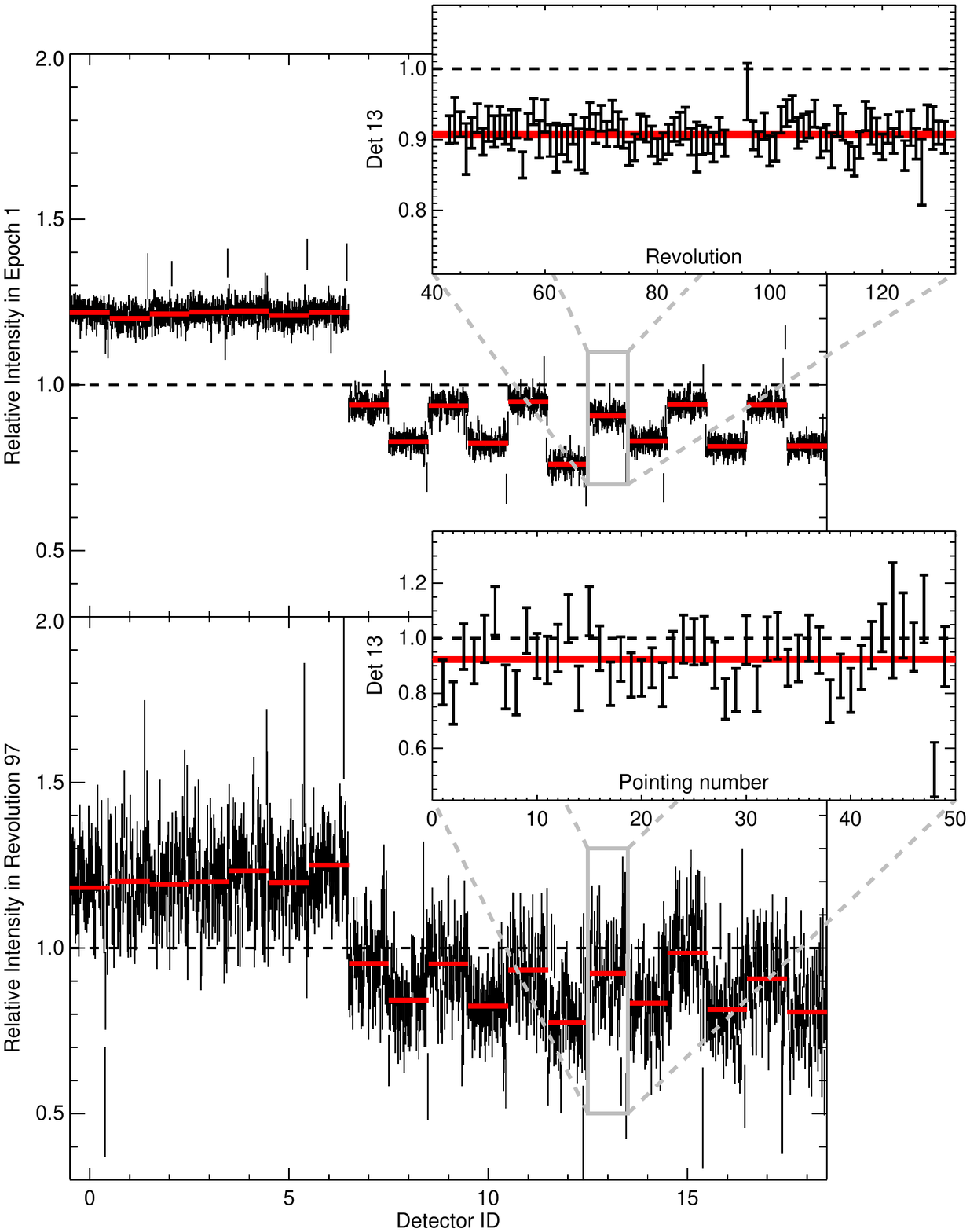}  
   \caption{Time variability of detector patterns. The detector pattern of the continuum (110-225 keV; \emph{above}, and a strong background line (1107 keV line; \emph{below}), illustrating variations on three characteristic time scales (epoch; \emph{(main graph; red bars in top panels)}, 3-day orbits \emph{(upper insert)}, and for one orbit no. 97 (red bars in lower graph) and pointings \emph{(lower inset)}.}
  \label{fig:SPI-detratio-times}
\end{figure}

\subsection{Background line categories}
Background lines result from de-excitation of a specific isotopic species, which are left behind in an excited state after interacting with a cosmic ray.
When the activated spacecraft materials are sufficiently distant from the camera, irradiation of the individual Ge detectors of the camera should be rather similar.
But upon closer inspection, there are deviations: some materials \emph{are} located in preferred directions relative to the Ge detectors, such as Ge nuclei themselves (camera centre versus periphery), or Bi from the BGO anticoincidence system (enclosing the Ge camera in all directions). 
The exposure to local background sources will therefore differ among detectors according to solid angle and shielding effects. Also the individual detector responses and efficiencies are not all identical, leading to significant differences.

Therefore we use as a characteristic the relative intensities  among the 19 different detectors of the Ge camera of SPI. We call this the \emph{detector ratio} parameter, or \emph{detector pattern}, defined as
\begin{equation}
r_j = N_{det}{{c_j t_j^{-1}}\over{\sum_n{c_n t_n^{-1}}}}
\label{eq:det-ratio}
\end{equation}
Here, $c_j$ are the event counts recorded for detector $j$ in an effective observing time interval $t_j$, and the factor $N_{det}$ of the number of detectors operating at the particular time normalises $r_j$ to 1.0 if all detectors record events at the same rate.

The equal-count ratio among detectors for isotropic detector and camera exposure to background appears to be approximately realised in continuum background. Its primary origin from cosmic ray activation translates here into processes distributed along the trajectory of the cosmic ray particle across the spacecraft and its environment. This is further smeared out by all secondary particles which may cause the photon background themselves: the electromagnetic cascade initiated by an incident cosmic-ray particle is expected to irradiate larger volumes of spacecraft and instrument, thus producing a diffuse particle and photon background originating from materials generally near the cosmic-ray trajectory. Different cosmic-ray trajectories then lead to an average origin of such continuum from the general spacecraft and instrument materials, { which leads to an} unspecific incidence direction with respect to SPI Ge detectors. The cosmic $\gamma$-ray continuum flux is orders of magnitude below such instrumental background. 
Fig.~\ref{fig:SPI-detratio-times} (upper graph) shows that indeed the detector ratio for continuum is { $\sim$1.0,} within a few percent deviations. Differences in actual volume of Ge material between detectors are even smaller and cannot account for these small differences. Some anisotropy from cosmic ray irradiation and from shielding effects, and the geometry of the spacecraft also contribute. Nevertheless, the near-equality of detector counts among all Ge detectors of the SPI camera for continuum also holds for different subsets of the data, as shown in 
Fig.~\ref{fig:SPI-detratio-times}: 
The inset expands the time axis for detector 13 to illustrate the variation over $\sim$90 orbits. The lower graph then shows the continuum background detector ratio for all detectors summed for one orbit, with the inset again expanding on the time scale to reveal variations among pointings, i.e. with $\sim$30~minutes resolution. 
These graphs illustrate that the detector ratio is rather well defined and constant over all time scales studied. Statistical variations are well within Poissonian expectations.

\begin{figure}
  \centering
  \includegraphics[width=0.8\linewidth]{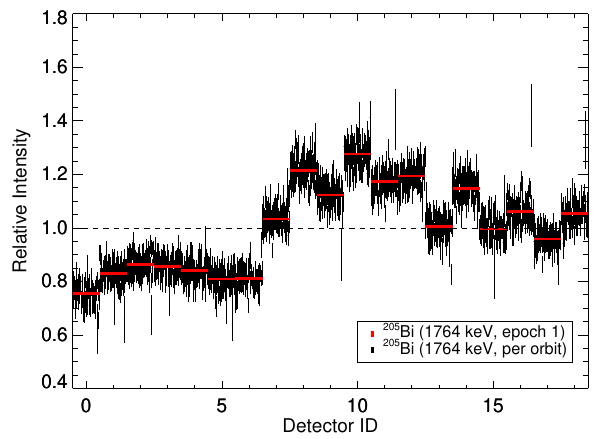}
   \includegraphics[width=0.8\linewidth]{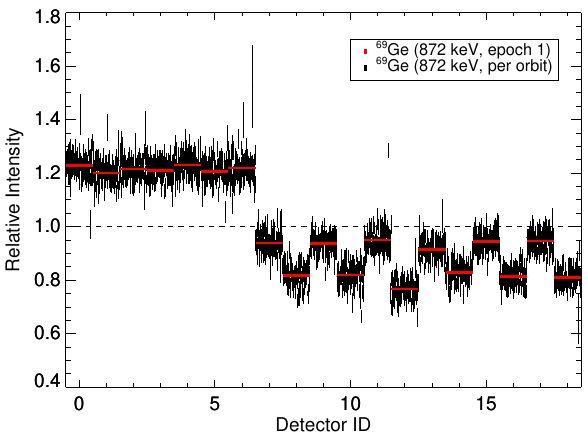}
     \includegraphics[width=0.8\linewidth]{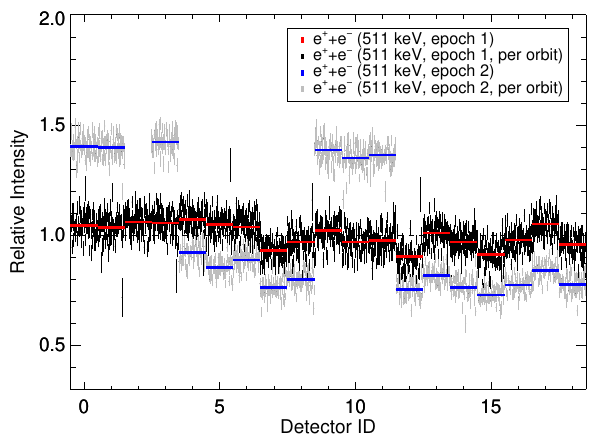}
  \caption{ Detector patterns for different types of background lines. { The} average of the first part of the mission with all 19 detectors working  \emph{('epoch 1') { is plotted as horizontal red lines}}, and for each orbit within this epoch { values per orbit  are plotted around this epoch-average value, using this} as a time axis within the epoch. {\it Above:} Line at 1764 keV from  $^{205}$Bi. {\it Middle:} Line at 872 keV from $^{69}$Ge. {\it Bottom:} Positron annihilation line at 511 keV, of mixed origins. For this, also shown is the situation where detector no.~2 had failed  (\emph{'epoch 2'; blue horizontal lines for epoch-average value, values per orbit superimposed { onto this and in grey}}).}
  \label{fig:SPI-detratio-lines}
\end{figure}

The lower part in Fig.~\ref{fig:SPI-detratio-times} shows the same detector ratio illustrations for one of the strongest lines attributed to Ge, at 1107 keV.
The detector-ratio pattern is very different from the continuum case above. Nevertheless, again a stable pattern is found over all time scales studied.
The Ge detectors on the outer edges of the Ge camera register significantly lower intensity (3 or 4 neighbouring detectors, respectively), while the inner camera detectors, which are surrounded by Ge material (6 neighbouring detectors), show this line at higher intensity. 
Also the $^{69}$Ge background line at 872 keV  (Fig.~\ref{fig:SPI-detratio-lines}) illustrates this same detector pattern, with differences of $\pm$30\% between detectors, all inner detectors having high intensities, while outer detectors show the lower background rates.
We see this pattern in all Ge related lines in a similar form, and also in lines from Zn, Ga, As, and Se, the secondaries from nuclear interactions with Ge, although differences in detail point to energy-dependent effects. 

In contrast, for Bi-isotope lines originating from the large BGO scintillator blocks that make up SPI's massive anticoincidence detector system, the inner detectors register a lower intensity:  they are partially shielded by the neighbouring detectors from the source in the anticoincidence detector system, in particular within the camera plane where those are closest (1764 keV line from $^{205}$Bi in Fig.~\ref{fig:SPI-detratio-lines}). Again, intensities vary $\pm$30\% between detectors. 

Another prominent characteristic pattern is found for the 511~keV line from positron annihilations. 
By nature, it receives contributions from a great variety of primary positron creation events, such as $\beta^+$ radioactivity and pair creation. Since the cosmic ray spectrum and intensity vary, different materials located at different distances to Ge detectors may contribute at different times. So similarly to the continuum, this also characterises a single narrow background line. Therefore this line may be expected to show more and different variability than any other background line. Indeed its detector pattern resembles more the one of continuum (Fig.~\ref{fig:SPI-detratio-lines}, bottom), but with a major change when detectors fail. 
When detector no.~2 had failed  (\emph{'epoch 2'} in Fig.~\ref{fig:SPI-detratio-lines}), multiple-detector hits involving the failed detector are falsely recorded as \emph{single-detector} events and so add to these genuine single-detector events. Therefore the relative intensity increases in detectors 0-1-3-9-10-11 which are located next to the failed one (see Fig.~\ref{fig:SPI-maskpattern}).

These examples illustrate that the detector-ratio pattern is helpful for a classification of background lines, as it encodes physical origin. 
We analysed all background lines fitted throughout the mission for such characteristic detector-ratio patterns  \citep[see also earlier work by][]{Augenstein:2015}. 
A cluster analysis of detector-ratio patterns was performed (see Appendix), and suggested the existence of
several useful categories:

\begin{itemize}
\item[(1)] 
 Lines with a rather flat pattern of detector ratios, similar to continuum. This includes internal transitions (IT), conversion X-ray photons (+L, +K), neutron- and proton-capture events, and also some light-element lines.
\item[(2)] 
 Ge-related lines. These show systematically-higher relative count rates in inner Ge detectors of the camera.    Lines most directly connected to Ge-detector events are prominent in that central detectors dominate strongly in the detector ratios. Three possible reactions are: electron captures on $^{69}$Ge, $^{68}$Ga, and $^{67}$Ga. They also include events, processes, and isotopes, related to Ge, Ga, As, Zn, Se, Mg, Al, and also some positron annihilation (when central detectors are somewhat enhanced).
\item[(3)] 
  Lines with a detector ratio pattern where outside detectors dominate. They may have origins in materials of the veto-shield and related isotopes, such as Bi, Pb, Ra, Rn, Tl, and Hg. Outer-detector dominance may be stronger or weaker. This includes also lines from natural radioactivities from the actinide alpha-decay-chains, i.e., Th, Fr, Ac, U. 
Processes involving materials of mountings and wirings, such as Co, Mn, Fe, Cu, Cr, V, and Ti fall into this category as well. The strong asymmetric detector ratio patterns, resulting, e.g., from IBIS instrument materials, such as Cd, Te, Cs, I, and neighbouring isotopes such as Xe or Ba are included also in this category. Some cases appear related to natural radioactivity of $^{40}$K.
   \item[(4)] 
 Lines with a weakly-pronounced, non-constant, but erratic detector-ratio pattern. These are often from Co or Cu, also neutron- and proton-capture reactions, and could be related to background events from the mask materials, such as W, Re, and Os. They may be also satellite lines or Compton edges of other lines. Lines with a strong erratic pattern could be unphysical as well, such as when used to represent broader spectral features.
\end{itemize}
\noindent
We use the results of the cluster analysis to further sub-divide the above four categories  where appropriate, and define seven categories of detector-ratio patterns (see Appendix, Table~\ref{table_SPI-line-table} column 5).

From their temporal intensity behaviour, we also can group the observed behaviour, and identify three categories:
\begin{itemize}
\item[(1)] 
 Lines clearly anti-correlated with the solar activity cycle. These follow in intensity the general continuum background. 
\item[(2)] 
 Lines with weak, if any, anti-correlation with the solar activity cycle. 
\item[(3)] 
 Lines where any anti-correlation with the solar activity cycle appears excluded, and sometimes may rather turn into a positive correlation with solar activity. 
\end{itemize}

\noindent
Table~\ref{table_SPI-line-table} in Appendix 1 includes this categorisation in two of its columns 5,6. More details are given there, both for the category definitions as well as for the individual lines.

%
\section{Modelling background}
\label{sec:background-modeling}

The behaviour of instrumental background as a function of pointing, detector, and energy bin can be modelled from the details extracted in our spectral fits (see Equ.~\ref{eq:model-fit}). 
But since these fits and the accumulated instrumental response data base were derived under the assumption of background dominating all data, the background model that will be derived in this way needs to allow for scaling of total intensity, to allow for the (small) celestial contribution to data. 

Separation of components of the background has been achieved in our spectral fits. Continuum, and a large number of lines, are each separated and tracked. Within the instrument, we thus expect that, while intensities may vary on various time scales, some general properties of the background components remain unchanged, or can be predicted.
The spacecraft orientation with respect to the background source, which is the spacecraft itself, is constant in time. In Figures.~\ref{fig:SPI-detratio-lines} and \ref{fig:SPI-detratio-times}, the constancy of detector ratios for different processes in the satellite is confirmed on all time scales. Therefore, statistical limitations in predicting background counts (see Fig.~\ref{fig:SPI-spectrum_all-mission}), can be reduced by determinations of the detector ratios from longer integration time scales. Detector patterns change slightly from energy bin to energy bin, as the degradation of the detectors distorts the spectrum in each detector somewhat differently. Our detailed spectral fits using Equ.~\ref{eq:continuum}  allow to determine these patterns, even in sub-resolution energy bins (here 0.5~keV) with great accuracy, as spectral response of each detector has been separated in these fits. The detector ratios in each energy bin, as constructed from the data base of spectral fit parameters, build the background model components $B_{jk}$ in Equ.~\ref{eq:model-fit}, separately for lines and continua in different energy bins $k$. As the relative contributions from lines and continuum changes with time (Fig.~\ref{fig:bgdvariations_LineContRatios}), an individual treatment, i.e. two distinct background detector patterns, is required.

The final step of the procedure then determins the intensities of different background components with their detector-ratio patterns, $\theta_j$ in Equ.~\ref{eq:model-fit}. This is done in a maximum likelihood fit, using Poissonian statistics. In particular, the log-likelihood of the Poisson statistics \citep[\emph{Cash statistics}, ][]{Cash:1979} is used. 
Sequential application of this procedure for the energy bins to be analysed will build the desired celestial spectrum, as well as the background spectrum, and their uncertainties. 

Background modelling requires specific adjustments for the physics within a spectral range of interest, such as the astrophysical $^{26}$Al, 511 keV, or $^{56}$Ni decay lines. A full and specific description of background modelling for such astrophysical cases is beyond the scope of this paper, and provided elsewhere \citep[][and Siegert et al., in preparation]{Siegert:2015,Diehl:2015,Siegert:2017,Siegert:2016}.

\section{Summary and Conclusions}
\label{sec:conclusions}

The INTEGRAL $\gamma$-ray spectrometer SPI \citep{Vedrenne:2003} has been in space since 2002, performing legacy science measurements while at the same time collecting a large dataset for characterising the instrument properties themselves in its space environment. 
In particular, a large set of instrumental background lines are continuously measured, and allow a calibration of the in-orbit spectral response, as well as a characterisation of the large background in this space environment which challenges and limits the sensitivity for measuring celestial sources. 
We analyse 13.5 years of data, to determine detailed spectral-response and background characteristics as they evolve over this time. This complements the characterisations obtained before launch \citep{Attie:2003} and early in the mission \citep{Roques:2003,Weidenspointner:2003}. 

We have developed a detailed parametrisation of SPI spectra in terms of a smooth underlying continuum and superimposed lines. 
We include in our description of instrumental lines an asymmetry term, which caters for the degrading spectral resolution in the harsh space environment with its intense bombardment by cosmic rays. 
Through regular annealing cycles, the SPI team achieves a recovery of spectral degradation roughly every six months. 
We separately fit spectra for each detector over the orbits of the mission, and thus determine parameters for each detector's spectral resolution, as well as for the variety of background components.

We find that the spectral resolution of our Ge detectors is between 2 and 3~keV at energies 300--2000~keV, and varies in a systematic saw-tooth pattern on a time scale of $\sim$6 months, with nearly linear degradation with time, and step restorations with the annealings. 
The annealings all were performed within the recoverable range of degradation. There is a small long-term trend of spectral resolution degradation of 0.3\% per year, in addition. The spectral resolution versus energy can be described by a square-root dependence, which is close to the pre-launch behaviour.

The energy calibration approach as adopted in the pre-processing step uses a set of six strong instrumental lines, readjusting their Gaussian peak values per each orbit. This incurs some variation of the absolute energy deposit precision, because the line shapes deviate from Gaussians due to the response degradations. Our spectral fits determine the absolute energies for all stronger lines to $\sim$0.03~keV, and they remain close to the pre-processing energy calibration, within a few tenths of a keV.  

Several strong components of background clearly anti-correlate with solar activity. This dominates the intensity changes in background. But in detail, the dynamic ranges and the reactions to particle irradiation events differs significantly between the general continuum and the various background lines. Lines can be grouped by physical-process similarities, and the relative intensity of identical background lines in different Ge detectors of the camera provides an excellent classification criterion towards this. Using in addition the frequently characteristic temporal variations of intensities, we can trace the physical origins of most instrumental lines, and provide classification of the lines into a set of categories distinguished by physical processes.

We 
provide the data and software tools for incorporation of such spectral-response and background knowledge into the scientific analysis  (see Appendix). 

%
\begin{acknowledgements}
 We are grateful to Astrid Augenstein for contributions during her Master thesis, which established initial versions of several methods and results of this paper. 
 { We thank the anonymous referee for constructive comments that helped to improve the paper.}
   The INTEGRAL/SPI project
  has been completed under the responsibility and leadership of CNES;
  we are grateful to ASI, CEA, CNES, DLR (No. 50OG 1101 and 1601), ESA, INTA, NASA and OSTC for
  support of this ESA space science mission.
\end{acknowledgements}
%
\bibliographystyle{aa}

%
\hfill\vfill\newpage

\appendix  
\section{The Spectral response and background data base - specification and access tools}	
We provide the full data base of spectral parameters for all single detectors (00-18) and all INTEGRAL revolutions between orbit numbers 43 and 1730 (with gaps due to annealing phases where no SPI data are taken) for energies between 20 and 2000~keV. This includes $N_{er}=39$ individual energy regions with interval sizes $\Delta E$ between 16 and 140 keV. From 20 to 1392~keV, and from 1745 to 2000~keV, single event data (SE) without pulse-shape identifiers are used (33 bands), whereas between 1392 and 1745~keV, pulse-shape-discriminated event data (PE) was used to suppress features in electronically affected spectral regions (6 bands\footnote{We also provide spectral fit results for the pulse-shape-discriminated event type in the entire range between 490 and 2000~keV, in addition to the single event type.}).
 
The spectrum in each energy band $i$ is characterised by a parametrised function, consisting of a continuum $(C_i(E)$, 2 parameters), and a set of distorted Gaussian lines $(L_i^l(E)$, 4 parameters each). The entire continuum is thus described as a multiple-broken power-law, each normalised to the centre of each individual energy band $(x_(c,i))$, see Equ.~\ref{eq:continuum}:

\begin{equation}
C_i(E) = C_{0,i} \left(\frac{E}{E_{m,i}}\right)^{\alpha_i}\mathrm{.}
\end{equation}

For lower energies, high count statistics allow us to determine each of the four line-shape parameters individually for each line, whereas at higher energies ($\gtrsim 1.7$~MeV), lines in a common energy band share the same degradation, in order to avoid over-parametrisation. For each line $l$, the line-shape is a convolution of a symmetric Gaussian $G_i^l(E)$, with an exponential tail function $T_i^l(E)$, accounting for the gradual degradation of the Ge detectors,
see Equ.~\ref{eq:line-function} and \ref{eq:gaussian}, \ref{eq:tail}.


Each of these parameter sets, $(C_0,\alpha,A_0^l,E_0^l,\sigma^l,\tau^l )_i$, in an energy band $i$, containing a number of lines $l$, is determined for each detector $j$, and each orbit $o$. For each detector, this amounts to $N_{er} \times 2=78$ fitted continuum parameters. The different energy regions contain different numbers of lines, depending on the suitability of the spectral region and the fit quality, ranging from $N_l=$2 to 27. In total, there are $N_l^{tot}=383$ identified lines, either as part of a complex line blend, or isolated, strong and narrow. This amounts to $N_l^{tot} \times 4=1532$ line-shape parameters per orbit and (working) detector, i.e. in total, $N_{par}^{spec}=1610$ spectral fit parameters, among which $N_{amp}=422$ are amplitudes, are determined for the $3960$ spectral energy bins. Over the { 13.5 year mission history, $N_{orb}=1556$ orbits were}  included in our data base, so that for a working detector, $N_{par}^{spec} \times N_{orb} = 2,505,160$ raw fit parameters have been determined. The total number of non-zero data base entries, i.e. excluding the failed detectors times, is then $40,412,610$.

The four line-shape parameters by themselves could provide specific information about the detector degradation and their intrinsic resolution, as they evolve with time. However, all parameters of a specific line are often degenerate, so that only combinations of parameters yield a physically meaningful measurement. 
The line width (FWHM) can be determined from a function of shape parameters $\sigma$ and $\tau$,

\begin{equation}
\Gamma_L = \Gamma \left[a_0+\sqrt{(1-a_0 )^2+((a_1 \tau)/\Gamma)^2 }\right]\mathrm{,}
\label{eq:fwhm_line}
\end{equation}

where $\Gamma= 2\sqrt{2 \ln(2) } \sigma$ is the FWHM width of the symmetric Gaussian $G(E)$, and $a_0=0.913735$, and $a_1=0.710648$ \citep{Kretschmer:2011} are numerical constants of this hyperbolic approximation. This Equ.~\ref{eq:fwhm_line} was used to evaluate the data base towards the spectral response as shown in Figures~\ref{fig:spectralResolution} to \ref{fig:recovery-dets}.

The peak position of the convolved line shape $L(E)$ is not simply given by parameter $E_0$, but rather has to be determined formally by

\begin{equation}
\left. \frac{\partial L(E)}{\partial E} \right|_{E=E_{peak}} \stackrel{!}{=} 0\mathrm{.}
\label{eq_epeak_formula}
\end{equation}

This reduces to $E_{peak} \simeq E_0 - \tau$ for small values of $\tau$.

The count rate $I$ of a line, in units of $\mathrm{ph~s^{-1}}$, is given as the integral over the energy of the full line, divided by the measurement time $T_{obs}$,

\begin{eqnarray}
I & = & \frac{1}{T_{obs}} \int_{-\infty}^{+\infty} L(E) dE = \frac{1}{T_{obs}} \int_{-\infty}^{+\infty} (G \otimes T)(E) dE = \nonumber \\
& = & \frac{1}{T_{obs}} \int_{-\infty}^{+\infty} G(E) dE = \frac{\sqrt{2 \pi}}{T_{obs}} A_0 \sigma\mathrm{.}
\label{eq_lintensity}
\end{eqnarray}

Note that the convolution preserves the area under the curve, so that the infinite integral of the convolved line shape is identical to the infinite integral of the symmetric Gaussian. Figures involving count rates, i.e. Figures.~\ref{fig:bgdvariations_rates}, \ref{fig:bgdvariations_LineContRatios}, \ref{fig:SPI-detratio-lines}, and \ref{fig:SPI-detratio-times}, have been evaluated using this Equ.~\ref{eq_lintensity}.

The data base of spectral response and background parameters is provided at 
http://www2011.mpe-garching.mpg.de/gamma/instruments/integral/spi/www/
in FITS file format. The files include the individual energy bands as separate extensions, and auxiliary information on how the fits were performed, e.g. which fitting functions were used, what parameter constraints have been set, or how many lines are included. In each FITS extension, {the number of rows equals the number of INTEGRAL orbits included in the data base times the number of detectors, i.e. for the 13.5 years $1556 \times 19=29564$}. There are 13 columns each, containing the revolution (1 column), the detector number (1) and the respective live-time (1), and the full list of fit parameters as itemised into continuum parameters (1) and line amplitudes (1), energies (1), widths (1), and degradations (1), as well as the respective uncertainties (5).

On the web page we also provide a detailed description on how the FITS files are structured, as well as tools, written in IDL, to calculate the physical parameters from the raw data base. These include the count rates for continua and lines, FWHM of the convolved line shapes and their peaks, and also the explicit form of the fitting function. In addition, the user may use a tool to query dates, detectors, and/or energies to obtain information about the SPI in-flight resolution, degradation or count rates, either for specific processes or entire energy intervals.

\begin{figure*}
  \centering
  \includegraphics[width=\linewidth]{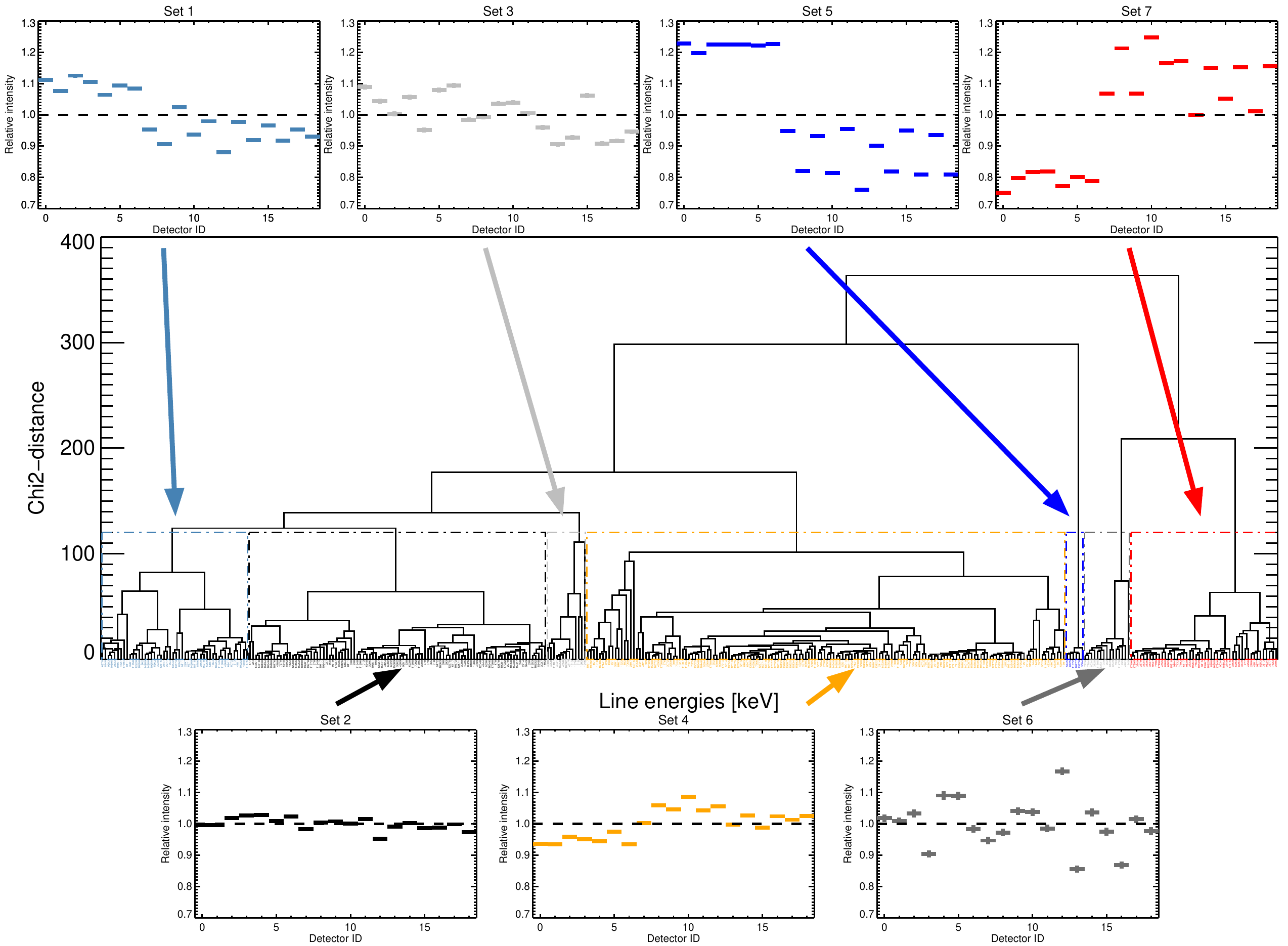}
    \caption{Cluster analysis for the detector-ratio patterns. The dendrogram of the clusters identified are shown in the main graph. The vertical axis reflects the differences between patterns, i.e., differences in the $\chi^2$ values. On the periphery, the detector-ratio graphs of the seven resulting pattern categories are illustrated.  }
  \label{fig:detratio_clusters}
\end{figure*}

\section{Background lines and their classifications}
Line features  can be identified in detector spectra down to low intensity levels if accumulation times are long enough (> days, ideally full mission). The lines identified in this way are listed in Table~\ref{table_SPI-line-table}. Herein we include all lines which may relate to a specific energy deposit in the detectors, i.e., all lines which are roughly consistent in width with instrumental resolution. We also included Gaussian lines in our spectral modelling which were used to model broader spectral features, for the sake of obtaining a satisfactory spectral model; these Gaussians are not included in this list, as physical origins are not sufficiently well related to a specific energy. In Table~\ref{table_SPI-line-table}, we provide additional information to each line entry, including classifications based on relative detector intensity ratio and intensity variation in time.

We performed a similarity analysis for the patterns in time and with detectors, as they are discussed in Section~5 above. The result for the detector patterns is shown in Fig.~\ref{fig:detratio_clusters}. Here, we use the differences between different patterns in their values of sqrt($\chi^2$) as a measure of similarity. Small differences within crowded areas define a cluster, and dissimilarity (large chi2 steps between isolated regions) defining the different clusters. The seven clearly identified clusters are marked in Fig.~\ref{fig:detratio_clusters} by coloured boxes, their detector pattern is shown in the plots at the circumference.

\noindent
\subsection{Further annotations to Table~\ref{table_SPI-line-table} : }\hfill\newline
\emph{Specifics to column 4:} '?' indicates an unknown process, '*' a significant change with respect to Weidenspointner et al. (2003), either due to change in process and/or energy, due to additional information, { 'CE xxx?' indicates possibility of the line being the Compton edge from the line at xxx~keV,} '**' indicates "satellite" or unphysical lines which are only used to fit the spectrum properly, and '***' marks lines which required multiple components to be fitted adequately, e.g. strong blends, the true line shape might be a composite of neighbouring lines.

\noindent
 \emph{Specifics to column 5:} 
 We identify 7 categories: (1) lines being related to Ge-detector events, processes, and isotopes, such as Ge, Ga, As, Zn, Se, Mg, Al, and also positron annihilation (central detectors dominate weakly); (2) lines showing a "strong flat" pattern - this includes internal transitions (IT), conversion X-ray photons (+L, +K), neutron- and proton-capture events, and also some light elements; (3) lines with a weak erratic pattern, often from Co or Cu, also neutron- and proton-capture reactions, could be related to background events from the mask materials, such as W, Re, and Os; may be also satellite lines or Compton edges; (4) lines related to processes involving materials of mountings and wirings, such as Co, Mn, Fe, Cu, Cr, V, and Ti (outside detectors dominate weakly) - also includes strong asymmetric patterns, e.g. from IBIS materials, such as Cd, Te, Cs, I, and neighbouring isotopes such as Xe or Ba; related sometimes to veto-shield materials (see category 7), and natural radioactivity $^{40}$K; (5) lines most directly connected to Ge-detector events (central detectors dominate strongly): only three possible reactions, i.e. electron captures on $^{69}$Ge, $^{68}$Ga, and $^{67}$Ga; (6) lines with a strong erratic pattern, probably unphysical in many cases, could also relate to neutron- and proton-capture events; (7) lines with origins in materials of the veto-shield and related isotopes, such as Bi, Pb, Ra, Rn, Tl, and Hg, includes also natural radioactivities from the actinide alpha-decay-chains, Th, Fr, Ac, U, \dots. The strong and narrow lines that are used to calibrate the spectral response (FWHM) are identified by their energy in { boldface} print. Lines used by ISDC to calibrate the energy are marked in \emph{italics}.
 
\noindent
\emph{Specifics to column (6):}  Correlations with the solar cycle, with (1) being strongly anti-correlated ({ correlation coefficient} $\rho < -0.6$), (2) being weakly anti-correlated ($-0.6 < \rho < -0.2$), and (3) being not correlated  ($\rho > -0.2$) { (}with a tendency to even positive correlations{ )}; (7) a comment column which identifies special behaviours such as (1) broad lines, (2) detector pattern strongly affected by detector failures, (3) rate strongly influences by solar flares, and (4) isotopes with a half-life time greater than the duration of one orbit (e.g. radio-active build-ups, or rates which lag behind the solar cycle).

\begin{table*}

\end{table*}
\end{document}